\documentclass{aa}  
\usepackage{natbib}
\bibpunct{(}{)}{;}{a}{}{,}
\usepackage{graphicx}
\usepackage{txfonts}
\begin{document} 
\title{
Short-term variability and mass loss in Be stars}
\subtitle
{
IV.  Two groups of closely spaced, approximately equidistant frequencies \\
in three decades of space photometry of $\nu$ Puppis (B7-8\,IIIe) 
\thanks
{
Based in part on data collected by the BRITE-Constellation satellite mission,
built, launched and operated thanks to support from the Austrian
Aeronautics and Space Agency and the University of Vienna, the
Canadian Space Agency (CSA), and the Foundation for Polish Science \&
Technology (FNiTP MNiSW) and National Science Centre (NCN). 
}
\thanks
{
Based in part on observations made with ESO Telescopes at the La Silla 
Paranal Observatory under programme ID 074.D-0240.
}
}
\author{D.\,Baade\inst{1}
\and
A.\,Pigulski\inst{2}
\and
Th.\,Rivinius\inst{3}
\and 
L.\,Wang\inst{4}
\and
Ch.\,Martayan\inst{3}
\and
G.\,Handler\inst{5}
\and
D.\,Panoglou\inst{6}
\and
A.C.\,Carciofi\inst{7}
\and
R.\,Kuschnig\inst{8, 13}
\and
A.\,Mehner\inst{3}
\and
A.F.J.\,Moffat\inst{9}
\and
H.\,Pablo\inst{10}
\and
S.M.\,Rucinski\inst{11}
\and
G.A.\,Wade\inst{12}
\and
W.W.\,Weiss\inst{13}
\and
K.\,Zwintz\inst{14}
}
\institute
{
European Organisation for Astronomical Research in the 
Southern Hemisphere (ESO), Karl-Schwarzschild-Str.\,2, 
85748 Garching b.\ M\"unchen, Germany;   
\email{dbaade@eso.org}
\and
Astronomical Institute, Wroc{\l}aw University, Kopernika 11, 
51-622 Wroc{\l}aw, Poland
\and
European Organisation for Astronomical Research in the 
Southern Hemisphere (ESO), Casilla 19001, Santiago 19, Chile 
\and
Center for High Angular Resolution Astronomy and Department of Physics 
and Astronomy, Georgia State University, P.O. Box 5060, Atlanta, 
GA 30302-5060, USA 
\and
Nicolaus Copernicus Astronomical Center, ul.\,Bartycka 18, 00-716 
Warsaw, Poland
\and
Observat\'orio Nacional, Rua General Jos\'e Cristino 77, S\~ao
Crist\'ov\~ao RJ-20921-400, Rio de Janeiro, Brazil
\and
Instituto de Astronomia, Geof\'isica e Ci{\^ e}ncias Atmosf\'ericas, 
Universidade de S{\~ a}o Paulo, Rua do Mat{\~ a}o 1226, 
Cidade Universit\'aria, 05508-900 S{\~ a}o Paulo, SP, Brazil
\and
Institut für Kommunikationsnetze und Satellitenkommunikation, Technical 
University Graz, Inffeldgasse 12, 8010 Graz, Austria
\and
D{\' e}partement de physique and Centre de Recherche en Astrophysique du 
Qu{\' e}bec (CRAQ), Universit{\' e} de Montr{\' e}al, C.P. 6128, 
Succ.\,Centre-Ville, Montr{\' e}al, Qu{\' e}bec, H3C 3J7, Canada 
\and
AAVSO Headquarters, 49 Bay State Rd., Cambridge, MA 02138, USA
\and
Department of Astronomy and Astrophysics, University of Toronto, 50 St.\,George
St., Toronto, Ontario, Canada M5S 3H4
\and
Department of Physics and Space Science, Royal Military College of Canada, PO
Box 17000, Stn Forces, Kingston, Ontario K7K 7B4, Canada
\and
Institute of Astrophysics, University of Vienna, Universit{\" a}tsring 1, 
1010 Vienna, Austria 
\and
Universit{\" a}t Innsbruck, Institut f{\" u}r Astro- und Teilchenphysik, 
Technikerstrasse 25, 6020 Innsbruck, Austria
}

\date{Received:  ; accepted:  }
 
\abstract 
{In early-type Be stars, groups of nonradial pulsation (NRP) modes
  with numerically related frequencies may be instrumental for the
  release of excess angular momentum through mass-ejection events.
  Difference and sum/harmonic frequencies often form additional
  groups.
}
{The purpose is to find out whether a similar frequency pattern occurs in 
the cooler third-magnitude B7-8\,IIIe shell star \object{$\nu$ Pup}.}
{Time-series analyses are performed of space photometry with
  BRITE-Constellation (2015, 2016/17, and 2017/18), SMEI (2003--2011),
  and Hipparcos (1989--1993).  Two {\it IUE} SWP and 27 optical
  echelle spectra spanning 20 years were retrieved from various
  archives.}
{The optical spectra exhibit no anomalies or well-defined
  variabilities.  A magnetic field was not detected.  All three
  photometry satellites recorded variability near 0.656\,c/d which is
  resolved into three features separated by $\sim$0.0021\,c/d.
  First harmonics form a second frequency group, also spaced by
  $\sim$0.0021\,c/d.  The frequency spacing is very nearly but not
  exactly equidistant.  Variability near 0.0021\,c/d was not detected.
  The long-term frequency stability could be used to derive
  meaningful constraints on the properties of a putative companion
  star.  The IUE spectra do not reveal the presence of a hot
  subluminous secondary.}
{ $\nu$\,Pup is another Be star exhibiting an NRP variability pattern with
  long-term constancy and underlining the importance of combination
  frequencies and frequency groups.  The star is a good target for
  efforts to identify an effectively single Be star.}
  

\titlerunning{Satellite photometry of Be star \object{$\nu$\,Pup}}
\authorrunning{D.\,Baade et al.}

   \maketitle
%

\section{Introduction}
\label{intro} 

When a rapidly rotating massive star evolves, it may have to solve a
problem when angular momentum is transported from the contracting core
to the outer layers \citep{2011A&A...527A..84K, 2013A&A...553A..25G}.
Apart from mixing, nonradial $g$-modes may be involved in the
angular-momentum transport process \citep[][and references
therein]{2013ASPC..479..319N}.  These stars may appear as Be stars
which are unique in that only Be stars seem to possess circumstellar
disks built from self-ejected matter \citep{2013A&ARv..21...69R}.  In
these little to moderately evolved stars, radiative winds do not
produce major mass-loss rates \citep{2014A&A...564A..70K}.  Broad
consensus exists that the near-critical rotation 
\citep{2005A&A...440..305F, 2012A&A...538A.110M,
  2013A&ARv..21...69R} is a necessary condition for the mass loss.
Numerous observational studies \citep{1998ASPC..135..343R,
  2009A&A...506...95H, 2013A&ARv..21...69R, 2017sbcs.conf..196B,
  2018A&A...610A..70B} suggest that much of the star-to-disk
mass-transfer process takes place in discrete events, which cover a
wide range in cadence and amplitude \citep{2018A&A...613A..70S,
  2018MNRAS.479.2909B}.  Although having the appearance of mass-loss
events, they may actually be more accurately described as
angular-momentum-loss events.  Angular-momentum-loss rates were
recently measured for a large sample of Be stars \citep{2018MNRAS.476.3555R}.

Evidence is broadly increasing \citep{1998ASPC..135..343R,
  2017arXiv170808413B} that multi-mode nonradial pulsation (NRP) plays
a central role in the conditioning of the stellar atmosphere for
outbursts.  Which specific conditions do lead to an outburst is
entirely unknown except that temporarily much increased NRP amplitudes
can have triggering power.  Interacting NRP modes may operate the
valves of the angular-momentum-loss process.
\citet{2017arXiv170808413B} have presented a scheme of nested clocks
with hierarchically decreasing frequencies, which all derive from
stellar pulsation frequencies and govern the opening and closing of
the angular-momentum-loss valves.  More massive stars may have to open
these valves more often and/or more widely because they evolve more
rapidly.  This mass-dependence of the activity seems confirmed by the
long-term photometry of large numbers of Be stars with a wide range of
spectral subclasses \citep[e.g.,][]{1998A&A...335..565H,
  2018MNRAS.479.2909B}.  Brightenings indicate ejections of gas, which
reprocesses the stellar light; in edge/equator-on systems such events
lead to fadings when the disk attenuates the stellar light
\citep{2014ApJ...785...12H}.

\begin{figure*}
\includegraphics[width=18.3cm,angle=-90]{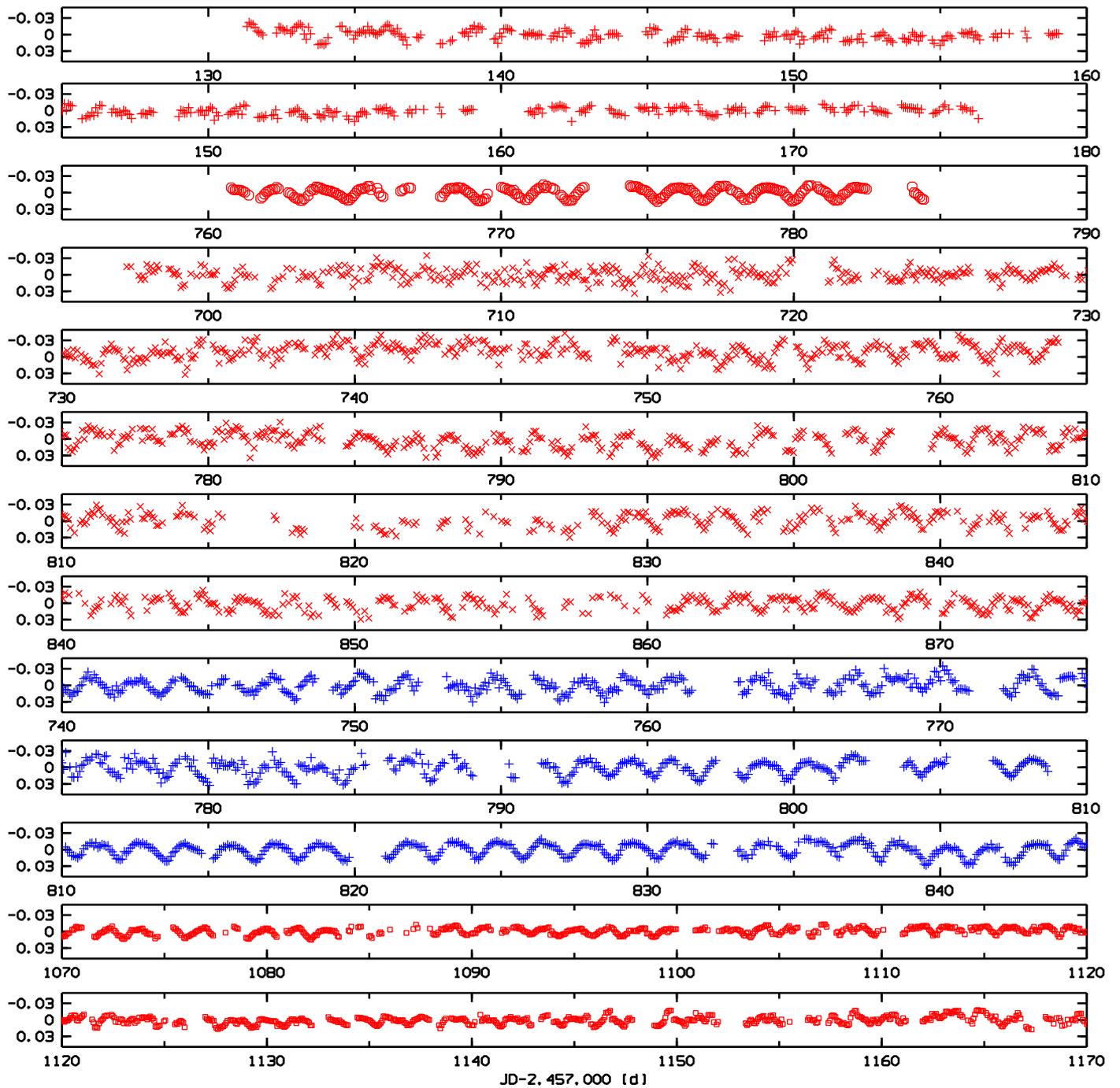}
\caption{All main strings of BRITE observations of $\nu$\,Pup in 2015,
  2017, and 2018.  Very long segments are split with partial
  duplications.  The ordinate is in units of instrumental magnitudes
  with arbitrary zero point.  Symbols: red $+$: BHr2015; red $\circ$:
  BHr2017; red $\times$: BTr2017; blue $+$: BLb2017; red $\square$:
  BHr2018.  Intercomparison of observations in identical time
  intervals but from different satellites shows that the most regular
  segments of the light curves are representative and the others
  deviate as a result of noise.  The amplitude was largest in 2017 and
  lowest in 2018.  In the bottom panel, between days 1145 and 1150 and
  shortly before day 1160, there may have been some mini outbursts. }
\label{LCmulti} 
\end{figure*}

\begin{figure*}
\includegraphics[width=5.9cm,angle=-90]{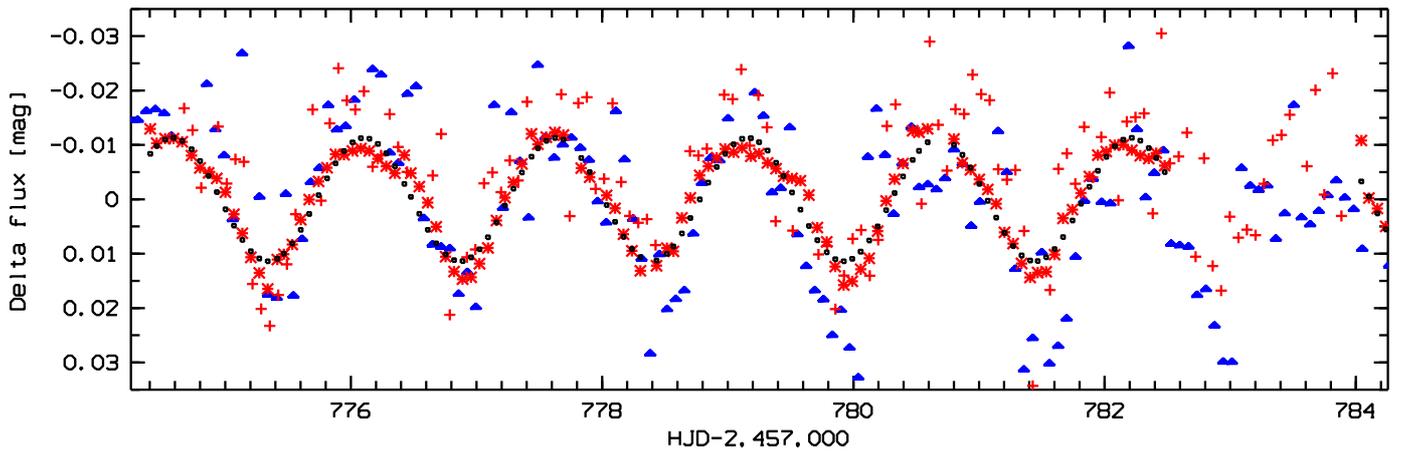}
\caption{A 10-d stretch (in 2017) of observations of $\nu$ Pup with
  BRITE satellites BLb (blue filled triangles), BHr (red asterisks),
  and BTr (red crosses).  The zeropoints of the magnitude scale are
  arbitrary.  Black dots represent a sine fit to the BHr data.  The
  deviation from sinusoidality is most clearly seen in the BHr data.}
\label{BLC} 
\end{figure*}

\begin{table*}
  \caption{Overview of the BRITE, Hipparcos, and SMEI observations.  
    Information not available for Hipparcos and SMEI is flagged as 'N/A'.
    Suffixes 'b' and 'r' in the 
    acronyms of BRITE satellites identify the spectral 
    passband (blue/red).  'CCDT' is the detector temperature range.  
    'Contig.\ time' denotes the typical time interval 
    per satellite orbit during which exposures were made. 'No.\ of TSA data 
    points' is the number of data points (in the case of BRITE formed as orbital averages 
    from the elementary measurements) and used for time-series analysis.
    Intervals between start and end dates include gaps.}
\label{obslog}
\centering
\begin{tabular}{l c c c c c c c c c}
\hline\hline
Satellite name      & P$_{\rm orbit}$  & Year(s)   &  JD--2,400,000     & CCDT             & Contig.\ time & No.\ of  & t$_{\rm exp}$ & No.\ of     & No.\ of TSA \\
(acronym)           & [min]          &           &  (start-end)       & [\degr C]        & [min]         & setups   &  [s]        & observ.      & data points \\
\hline
Hipparcos           & 637.2          & 1989-1993 &   47860-49051      &       N/A        &    N/A        &   N/A    &  $\sim$20   &    115      &     115     \\
SMEI                & 101.5          & 2003-2011 &   52673-55833      &       N/A        &    N/A        &   N/A    &  $\sim$900  &  38617      &   31707     \\
BRITE-Heweliusz     & 97.1           & 2015      &   57098-57176      &      9.3-18.8    &  2.6-29.8     &    3     &  1          &  47816      &     487     \\
(BHr)               &                & 2017      &   57760-57784      &     13.3-22.2    &  4.7-20.1     &    1     &  4          &  15542      &     257     \\
                    &                & 2017/18   &   58066-58222      &      0.9-23.8    &  3.0-21.4     &    7     &  4          &  63886      &    1433     \\
BRITE-Lem           & 99.6           & 2016/17   &   57738-57910      &     18.2-34.6    &  4.0-13.9     &    5     &  2          &  82516      &    1382     \\
(BLb)               &                &           &                    &                  &               &          &             &             &             \\
BRITE-Toronto       & 98.2           & 2016/17   &   57697-57877      &      2.9-20.4    &  3.7-15.7     &    6     &  4          &  65260      &    1822     \\
(BTr)               &                &           &                    &                  &               &          &             &             &             \\
\hline
\end{tabular}
\end{table*}

At the cool limit of the so-called Be phenomenon, spectroscopic
evidence of a disk is often restricted to shell absorption lines.
However, recent observations by \citet{2018A&A...609A.108S} have
demonstrated that, with the help of high-quality spectral line
profiles, weak H$\alpha$ emission can be detected in many more late
B-type stars than previously thought.  The variability of their
emission lines seems to be much slower and reach lower amplitudes than
displayed by many early-type Be stars \citep{2013A&A...556A..81B}.
Photospheric line-profile variability is also much less pronounced in
late-type Be stars \citep{1989A&A...222..200B}.

With the advent of powerful space photometers, a substantially more
detailed view of the variability of late-type Be stars is emerging.
Recently, {\it Kepler} observations of \object{KIC\,11971405}
(HD\,186567) \citep[B5\,IV-Ve;][]{2017A&A...598A..74P} revealed one of
the most intricate networks of NRPs known in Be stars.  Most
variabilities have sub-mmag amplitudes.  At B5, HD\,186567 is also one
of the cooler Be stars with detected NRPs.  Still cooler is
\object{$\beta$\,CMi} at B8, for which {\it MOST} recorded many NRP
modes with similarly low amplitudes \citep{2007ApJ...654..544S}.  In
observations of the B7\,V star \object{CoRoT 101486436},
\citet{2018A&A...613A..70S} find this star's variability to resemble
that of $\beta$\,CMi; however, the peak-to-valley amplitude of a
0.03\,c/d beat (perhaps: difference) frequency has an amplitude of
12,000\,ppm with individual frequencies near 5\,c/d reaching up to
3,000\,ppm.

During the monitoring period with {\it Kepler}, also KIC\,11971405
underwent some small outbursts.  As \citet{2017arXiv170808413B}
showed, the spacing in time of some of these events can be related to
difference frequencies of some NRP modes with the highest amplitudes.
Such coupling had previously been found in BRITE, {\it Kepler}, and
CoRoT observations of several early-type Be stars \citep[][Rivinius et
al., in prep.]{2016A&A...588A..56B, 2016A&A...593A.106R,
  2017arXiv170808413B}.  \citet{2017MNRAS.471.2882W} conclude that
{\it Kepler}/K2 observations of four Be stars in the Pleiades are
consistent with this hypothesis for the two least evolved Be stars,
\object{Merope} (B6\,IVe) and \object{Pleione} (B8\,Vne).

Late-B spectral types are also the realm of the historically mythical
Maia variables.  Maia itself (\object{20\,Tau}, B7\,III) was finally
shown not to be pulsating but exhibiting slow rotational modulation
\citep{2017MNRAS.471.2882W}.  However, there do exist probable
$g$-mode pulsators between the conventional Slowly Pulsating B Star
(SPB) and $\gamma$ Dor domains \citep[e.g.][]{2013A&A...554A.108M};
this includes late B-type stars.  Whether the pulsations of the other
six B stars in the Pleiades studied by \citet{2017MNRAS.471.2882W}
fall into just one single category is unknown.  The light curves of
\object{Alcyone} (B7\,IIIe), \object{Electra} (B6\,IIIe),
\object{Merope} (B6\,IVe), and \object{Pleione} (B8\,Vne) exhibit
clear variations of their upper and lower envelopes which may be due
to the beating of a few single frequencies.  A similar pattern is
absent in the normal binary B star \object{Atlas}
(B8\,III\,$+$\,B8\,V) and much less prominent in the other normal
B-type star Taygeta (B6\,IV) if present at all.
\citet{2017MNRAS.471.2882W} explicitly report frequency groups for the
Be stars Alcyone, Electra, and \object{Taygeta} and describe the
variability of Merope (Be) and Taygeta (B) in terms of closely spaced
frequencies, which is consistent with the presence of frequency
groups.

Frequency groups are characteristic of Be stars
\citep{2011MNRAS.413.2403B, 2017sbcs.conf..196B, 2018A&A...613A..70S}
but do not seem limited to them \citep{2015MNRAS.450.3015K}, and not
all pulsating Be stars display frequency groups.  The latter authors
proposed that frequency groups may be understood as clusters of
combination frequencies of NRP modes.  This was confirmed by
\citet{2017A&A...598A..74P} in the same data.  In the early-type Be
star \object{25\,Ori}, \citet[][an extended version of this work with
additional BRITE and revised SMEI observations is in
preparation]{2017arXiv170808413B} tentatively identified an extremely
rich pattern of grouped combination frequencies.

It has been proposed \citep{2008ApJ...685..489C, 2016A&A...593A.106R,
  2017MNRAS.471.2882W} that Be stars may be rapidly rotating SPB
stars.  However, among the B-type stars without Be characteristics,
neither SPB stars nor any other type of pulsating star span such a
wide range in effective temperature as the Be phenomenon does.  On the
other hand, pulsations of Be stars do not seem to significantly extend
to lower or higher temperatures than other pulsating OB stars do, and
the combined role of rapid rotation and mass and angular-momentum
leakage is still to be explored in detail.

The third-magnitude late-type Be star \object{$\nu$\,Pup} (HR\,2451,
HIP\,31685, HD\,47670) has not been given much attention by
spectroscopists.  \citet{1909ApJ....29..232C} described H$\gamma$ as
consisting of a broad absorption with superimposed fairly sharp
central absorption.  Such shell components occur in Be stars if the
star is viewed equator-on and the line of sight passes through the
disk.  In eight spectra, the radial velocity of the narrow component
varied from $+$33\,km/s in 1904 to $+$20\,km/s in 1908.  Temporary
shell absorption flanked by weak line emission in H$\alpha$ was also
detected by \citet{1999A&A...348..831R, 2006A&A...459..137R}.  Shell
absorptions were not reported by \citet{1897ApJ.....6..349P} and by
\citet{1975ApJS...29..137S} so that the disk is probably not
persistent.

\citet{1987SAAOC..11...93C} treated $\nu$\,Pup as a secondary
photometric standard.  On the other hand, from Hipparcos observations,
\citet{2002MNRAS.331...45K} reported a variability with frequency
0.15292\,c/d and amplitude 0.00117\,mag. At other wavelengths,
$\nu$\,Pup has remained inconspicuous: It was not detected in the
ROSAT All Sky X-ray Survey \citep{1996A&AS..118..481B}, and
\cite{2018arXiv180601294B} did not find it embedded in a mid-IR
nebulosity.  \citet{2008MNRAS.389..869E} do not list $\nu$\,Pup as
multiple.

$\nu$\,Pup was selected from the Be stars observed with
BRITE-Constellation because of its relatively low temperature
(extending earlier studies with BRITE of hotter Be stars), its
brightness (yielding a better signal-to-noise ratio with BRITE), and
the, at a first glance, unusual simplicity of its frequency spectrum.

\section{Observations} 
\label{observations}

New observations were obtained with the BRITE-Constellation of
nanosatellites.  Its mission and operations concepts were described by
\citet{2014PASP..126..573W}, who also provide the definition of the
BRITE-Constellation photometric passbands, and
\citet{2016PASP..128l5001P}, respectively.  A total of three of the
five cubesats were employed for different time intervals in 2015
(BRITE field 08-VelPic-I-2015), 2016/17 (BRITE field
23-VelPic-II-2016), and 2017/18 (BRITE field 33-VelPic-III-2017) as
detailed in Table\,\ref{obslog}, which also contains the number of raw
observations and derived data points. The BRITE data are available
from the BRITE photometry
wiki\footnote{http://brite.craq-astro.ca/doku.php}.  For simplicity,
the three datasets will below be referred to as B2015, B2017, and
B2018, respectively.  The rationale for additional processing of the
pipeline-reduced data \citep{2017A&A...605A..26P} was explained in
much detail by \citet{2018arXiv180209021P} and
\citet{2018arXiv180108496P}.  For the observations of $\nu$\,Pup, it
was implemented and applied as described in
\citet{2018A&A...610A..70B}.

Pipeline-reduced BRITE datasets are subdivided into so-called setups
if some observing condition changed during the observing season.  The
most common reason is re-acquisition of the field.  For all three
satellites, temperature jumps occurred within some setups, and the
correlations with CCD temperature of magnitudes before and after such
events were different.  Therefore such setups were split, and the
decorrelation was performed separately for each data segment.  An
overview of all major segments of BRITE observations can be gained
from Fig.\,\ref{LCmulti}.  Figure\,\ref{BLC} presents a ten-day
interval at higher resolution with observations from BLb, BHr, and BTr
in 2017.

Archival observations (Jackson 2018, private communication) with the
Solar Mass Ejection Imager \citep[SMEI;][]{2004SoPh..225..177J} extend
from Feb.\,2, 2003 through Sept.\,28, 2011.  Contrary to an earlier
study in this series using SMEI data \citep[][who also describe SMEI
from the stellar-photometry point of view]{2018A&A...610A..70B}, the
data now available identify with which of the three cameras a given
measurement was made. This vastly improves the scope of the
preparation of the data for time series analysis.

SMEI photometry exhibits large-amplitude annual variations the details
of which are strongly dependent on position in the sky because the
objective of SMEI was to observe light scattered by electrons ejected
by the Sun.  For $\nu$\,Pup, these artifacts were much reduced in a
multi-step procedure applied separately to the data from each camera.
First, for each data point its difference from a sliding average over
10 days was calculated in units of flux as well as of the standard
deviation in the 10-day bin concerned.  In a diagram with these two
parameters, several narrow rays can be seen fanning out from the
origin.  Outliers were clipped in both coordinates.  Second, a
second-order polynomial was fitted and subtracted, and newly apparent
outliers were eliminated.  In the third step, annual and half-annual
variations were removed by subtracting fitted sine functions.  After
folding the data with a period of one year, new outliers were
identified by eye (because of their strongly non-Gaussian
distribution) and removed.  These latter two steps were repeated for
frequencies between 0.02 and 0.06\,c/d, which may be (partly) related
to the Moon.  In the sixth step, the very pronounced 1-c/d variability
was subtracted, and, finally, remaining data points with large values
of SMEI parameter sigPSF were also removed. Thereafter, the three
datasets were merged for the subsequent time-series analysis.  All
together, these seven steps reduced the numbers of data points from
the initial 7127/18516/12974 for Cameras 1/2/3 to the final
5353/16380/9974, respectively.

A further archival dataset is available from Hipparcos
\citep{1997yCat.1239....0E}.  It consists of 115 photometric
measurements obtained between Nov.\ 30, 1989 and March 5, 1993.  There
are variations on timescales of several hundred days, and several of
them fold the data to well-behaved light curves.  They were not
removed because none of them stands out above the others.  The dataset
contains only one apparent outlier, which was retained.

\begin{figure}
\includegraphics[width=4.65cm,angle=-90]{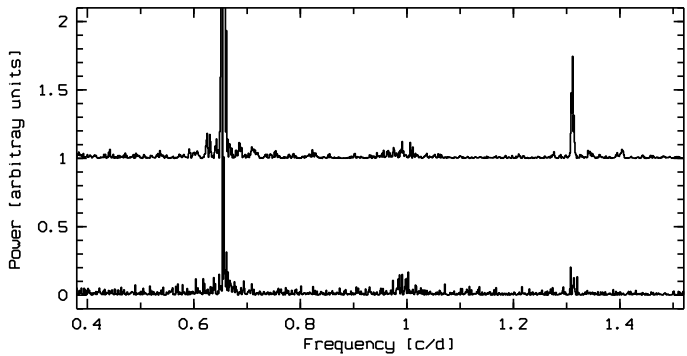}
\caption{Power spectra of BRITE (vertically offset for clarity; all
  datasets combined) and SMEI (bottom) photometry.  Two
  frequency groups near 0.656\,c/d and 1.31\,c/d as well as 1-c/d
  variability are present.  The power spectra were scaled to the same
  peak strength of the 0.656-\,c/d groups.}
\label{PS} 
\end{figure}

The time-series analysis (TSA) of the BRITE, SMEI, and Hipparcos data
was performed with implementations of Scargle's method
\citep{1982ApJ...263..835S} and plain power-spectrum calculation in
the TSA package of MIDAS \citep{2003ASSL..285...89B}.  The differences
between the methods seem negligible.  With orbital periods near
100\,minutes, both BRITE and SMEI have Nyquist frequencies around
7\,c/d.  The final power spectra between 0.4\,c/d and 1.5\,c/d for
SMEI and all BRITE data are compiled in Fig.\,\ref{PS}.

\begin{table}
  \caption{Appearance of the H$\alpha$ line in \object{$\nu$\,Pup} at various 
    times.  Significant variations in the H$\alpha$ profile 
    were only seen between the time intervals listed, not within any of them.
    Notations:  n = number of spectra, E = emission, S = shell absorption.  
    Julian Dates of the individual spectra (except for ESPaDOns) are presented 
    in Table\,\ref{spectra}.}
\label{Halpha}
\centering
\begin{tabular}{l c c l}
\hline\hline
Time & Spectrogr. & n & Appearance of H$\alpha$ \\
\hline
1995 May     & {\sc Heros} &  1  & very weak E, clear S  \\
1996 Mar     & {\sc Heros} &  2  & very weak E, clear S  \\
1996 May-Jun & {\sc Heros} &  6  & as in 1996 Mar        \\
1997 Jan-Apr & {\sc Heros} &  7  & extrem.\ weak E,      \\
             &             &     & weakened S            \\
1999 Jan     & FEROS       &  1  & extremely weak E,     \\
             &             &     & very weak S           \\
1999 May-Jun & {\sc Heros} &  5  & almost no E,          \\
             &             &     & very weak S           \\
1999 Jul     & FEROS       &  2  & E and S weaker than   \\
             &             &     & in 1999 Jan           \\
2000 Jan     & FEROS       &  1  & no E, no S            \\
2005 Jan     & FEROS       &  3  & no E, no S            \\
2016 Jan     & ESPaDOns    & 40  & no E, no S            \\
\hline
\end{tabular}
\end{table}

Pipeline-reduced spectra were downloaded from the {\sc
  Heros}\footnote{https://www.lsw.uni-heidelberg.de/projects/instrumentation/\newline
  Heros/search.html}, the ESO\footnote{archive.eso.org} (for FEROS),
the
CFHT\footnote{http://www.cadc-ccda.hia-iha.nrc-cnrc.gc.ca/en/cfht/}
(for ESPaDOns), and MAST\footnote{https://archive.stsci.edu/iue/} (for
IUE) online archives.  {\sc Heros} \citep{1998RvMA...11..177K}, FEROS
\citep{1999ASPC..188..331S}, and ESPaDOns \citep{2006ASPC..358..362D}
are optical echelle spectrographs with approximate resolving powers of
20,000, 48,000, and 68,000, respectively.  With {\sc Heros} and FEROS
only single spectra per night were obtained every couple of days or
weeks \citep[they include the spectra used by][]{2006A&A...459..137R}.
By contrast, all 40 ESPaDOns spectra were secured within about one
hour.  A description of the {\it International Ultraviolet Explorer
  (IUE)} is available from \citet{1978Natur.275..372B}.

\section{Analysis}
This section is structured as follows: Sect.\,\ref{spectroscopy}
describes the appearance of the spectra; a simple stellar model is
provided, radial velocities are measured and analyzed for periodic
variations, and limits are placed on the UV flux contributed by a
sub\-luminous hot companion.  A general overview of the photometric
variability is given in Sect.\,\ref{photoover}.  It is followed by the
description of the two frequency groups found (Sect.\,\ref{coarse})
and their simulation (Sect.\,\ref{simulations}).  The final
subsections deal with mass loss (Sect.\,\ref{massloss}) and, from
photometric Doppler shifts, derive limits on the mass and orbital
period of a putative companion (Sect.\,\ref{Doppler}).

\subsection{Spectroscopy}
\label{spectroscopy}

In the ESPaDOnS spectra, the strengths of HeI\,4471 and MgII\,4481 are
indistinguishable.  In the classification scheme of
\citet{1968ApJS...17..371L}, this corresponds to spectral type B8 on
the main sequence.  \citet{1897ApJ.....6..349P} and
\citet{1975ApJS...29..137S} classified $\nu$\,Pup as B8.  However, the
weakly developed wings of the Balmer lines indicate luminosity class
III-IV for which case \citet{1968ApJS...17..371L} notes that the He
lines are weaker than on the main sequence.  This suggests an MK type
closer to B7\,III.  For B7\,III and B8\,III, the grid of stellar
parameters calculated by \citet{2018A&A...609A.108S}, which is based
on \citet{2009A&A...501..297Z}, is provided in
Table\,\ref{stellarPar}.  The derived parameter values are comparable
to those compiled by \citet{2017MNRAS.471.2882W} for the B-type stars
in the Pleiades.  However, the calculated $v$\,$\sin$\,$i$ of
170\,km/s is not a good match to the observations which are in much
better agreement with the value of 246\,km/s listed by
\citet{2002A&A...381..105R}.  In the following, averages of the two
parameter sets are adopted.

Because of the high apparent brightness and late spectral sub-type of
$\nu$\,Pup, the star has a relatively large parallax.  The revised
Hipparcos measurement of 7.71\,mas \citep{1997A&A...323L..49P,
  2007A&A...474..653V} of 8.78\,mas is in excellent agreement with the
Gaia DR2 value of 8.9182\,mas \citep{2018yCat.1345....0G}.  However,
for both projects, stars as bright as $\nu$\,Pup are a challenge.
With 0.75\,mag as the bolometric correction and 0.05\,mag for the
reddening, the averaged parallax of 8.85\,mas implies a bolometric
magnitude of -2.9\,mag, 0.26\,mag brighter than the average of the two
models (Table\,\ref{stellarPar}).  The difference is not significant;
however, if it is attributed entirely to the radius estimate (for a
uniform stellar disk), the radius derived from the observations is
12\% larger than the calculated one.

For the time spanned by the spectra, the evolution of the H$\alpha$
line is documented in Table\,\ref{Halpha}.  A very weakly developed
circumstellar disk was directly detected only between 1995 and 1999.
However, the presence of strong central quasi-emission features in
HeI\,4471 in 1985 \citep{1989A&AS...79..423B} suggests
\citep{1999A&A...348..831R} that the disk may have been much denser a
decade before.  Perhaps, considering the general slow variability of
disks around late-type Be stars, the 1995-1999 observations only trace
the terminal phase of the disk dissipation.

\begin{table}
  \caption{Stellar parameters calculated for spectral types B7\,III 
    and B8\,III as described in \citet{2018A&A...609A.108S}.  Colons (:) 
mark uncertain values.}
\label{stellarPar}
\centering
\begin{tabular}{l c c c}
\hline\hline
Parameter                 &    Unit      & B7\,III    & B8\,III   \\
\hline
Equatorial radius         & R$_\odot$     &     7.1    &     6.8   \\
Polar radius              & R$_\odot$     &     6.1    &     5.8   \\
Mass                      & M$_\odot$     &     3.5    &     3.1   \\
log ($L$/L$_\odot$)        & N/A          &     3.0    &     2.9   \\
T$_{\rm equ}$               & K            &  10900:    &  10020:   \\
T$_{\rm pole}$              & K            &   14200    &   13060   \\
T$_{\rm eff}$               & K            &   12790    &   11760   \\
log ($g_{\rm equ}$/[cgs])   & N/A          &     3.0    &     2.9   \\
log $g_{\rm pole}$          & N/A          &     3.4    &     3.4   \\
$v$\,$\sin$\,$i$ (adopted)& km/s         &    170     &    170    \\
$v_{\rm equ}$               & km/s         &    179     &    173    \\
$v_{\rm Kepler}$             & km/s         &    305     &   294     \\
f$_{\rm rot}$               & c/d          &    0.5     &    0.5    \\
Fract.\ crit.\ rot.\ rate & N/A          &  0.587     &  0.587    \\
(fixed)                   &              &            &           \\
\hline
\end{tabular}
\end{table}

All stronger non-Balmer lines were very carefully scrutinized for
line-profile variations.  However, because the depth of the lines is
only about 5\% of the adjacent continuum level, non-Gaussian pixel
noise and instrumental variations in the continuum shape make this
search exceedingly difficult.  The width of the best isolated of these
lines, HeI\,4026, measured as the full width at half maximum of a
fitted Gaussian, was constant to $\pm$4.1\% in the {\sc Heros} and to
$\pm$2.5\% in the FEROS spectra.  This scatter is plausibly explained
by the said imperfections of the spectra.  The variability is stronger
in the line wings.  In some of the best defined line profiles, the
wings assume the shape of a ramp on either the blue or the red side,
while the respective other wing is much steeper.  This can be the
signature of nonradial pulsation \citep{2003A&A...411..229R}.  Phasing
of selected line profiles with the main photometric frequency near
0.656\,c/d (cf.\ Sect.\,\ref{photoover}) did not bring out a clearly
coherent pattern because instrumental properties dominated.  No
obvious variations were seen in the strength of any line, in agreement
with earlier observations by \citet{1989A&AS...79..423B}, who obtained
two pairs of high-resolution, high-signal-to-noise profiles of
He\,I\,4471 and Mg\,II\,4481 in two consecutive nights.

\begin{figure}
\includegraphics[width=6.5cm,angle=-90]{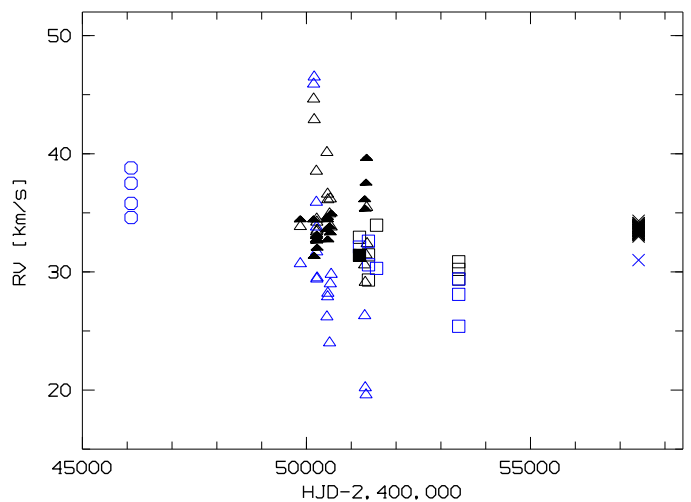}
\caption{Radial velocities of stellar H$\delta$-H8 lines (open black
  symbols), other stellar lines (He\,I\,4026.2, He\,I\,4143.8,
  He\,I\,4471.5, and Mg\,II\,4481.2; open blue symbols), and the
  circumstellar shell line in H$\alpha$ (filled symbols).  It can be
  seen that the accuracy of the ESPaDOnS measurements (crosses near
  mJD\,57500) is highest, that of the {\sc Heros} measurements
  (triangles) lowest.  Measurements with FEROS (CES) are plotted as
  squares (circles).  See also Table\,\ref{spectra} and
  Sect.\,\ref{spectroscopy}.}
\label{BalmerRV} 
\end{figure}

Radial velocities (RVs) were measured in the H$\beta$-H12 Balmer,
several He\,I, and some metal lines.  The results are given in
Table\,\ref{spectra} and Fig.\,\ref{BalmerRV} separately for (i) the
H$\gamma$-H8 Balmer lines (which are unaffected by circumstellar
components), (ii) HeI $\lambda\lambda$ 4026, 4143, and
4471 as well as MgII\,4481 and, when possible, for the shell
absorption component in H$\alpha$.  The measurements in the Balmer
lines are probably compromised by the residual echelle ripple in the
pipeline-reduced spectra.  The radial velocities of the other stellar
lines suffer from this, too, but mainly from pixel noise; there may
also be an effect of the suspected line-profile variability.  The
homogeneity of the shell velocities is reduced by variable telluric
absorption features.

\begin{figure}
\includegraphics[width=7.1cm,angle=-90]{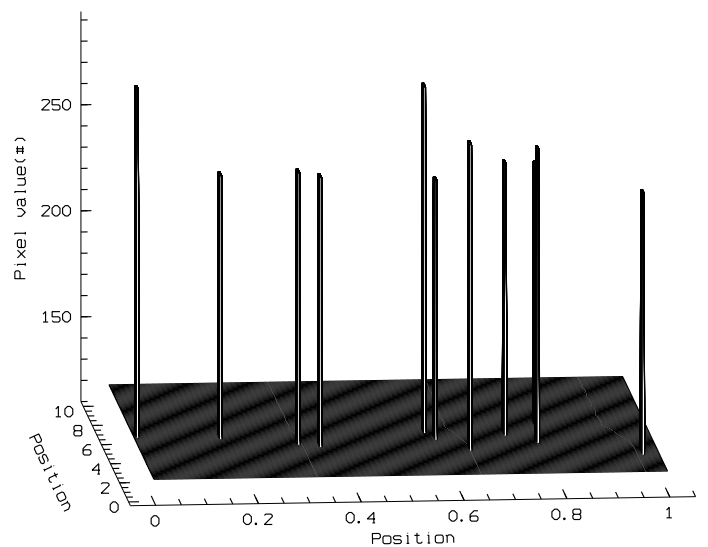}
\caption{Plane with coordinates of frequency (in c/d, abscissa) and
  radial-velocity amplitude (in km/s, ordinate).  The data axis
  provides the number of frequency-amplitude data pairs resulting from
  circular orbits fitted to the radial velocities in
  Table\,\ref{spectra}, as described in Sect.\,\ref{spectroscopy}.  Each
  spike has a footprint of dimensions 0.0008\,c/d and 0.1\,km/s;
  only for better visibility were they enlarged.  The floor level is
  $\leq$30$\pm$6 data points per pixel. }
\label{AmpFrqMap} 
\end{figure}

Because the extremely uneven distribution of epochs of the RV
measurements (Table\,\ref{spectra}, Fig.\,\ref{BalmerRV}) prevents the
direct determination of a putative orbital period, the full parameter
space was explored to find out where any permissible solutions might
occur.  Frequencies and RV amplitudes were derived simultaneously from
sine fits to the non-Balmer RVs in Table\,\ref{spectra} with 250,000
equally spaced start frequencies from 0.000004 to 1\,c/d.  This
sampling was chosen to match the time span of the spectra of 11,300\,d
(Table\,\ref{spectra}).  The mean amplitude is 2.95$\pm$1.3\,km/s with
a maximum of 8.5\,km/s so that any real circular orbital amplitude
should not exceed 10\,km/s.  In Fig.\,\ref{AmpFrqMap}, the results are
mapped onto the frequency-amplitude plane binned to pixels of
0.0008\,c/d and 0.1\,km/s, respectively.  On average, the number of
frequency/amplitude pairs per such bin is 10.  However, the number of
amplitudes above 4\,km/s is very small so that 30$\pm$6 pairs per
pixel is a more realistic estimate of the floor value.  The peaks in
Fig.\,\ref{AmpFrqMap} reaching levels of 250 pairs are formally highly
significant.  However, in a genuine binary system, only one such pixel
can represent the correct frequency and amplitude.  Since no peak
stands out above the others, they only permit an upper limit to be
placed on the amplitude of a putative binary.  Inspection of
Fig.\,\ref{AmpFrqMap} supports the validity of the above value of
10\,km/s.  By contrast, the possible period is not usefully
constrained by the RVs.

The ESPaDOnS observations form part of the BRITE spectropolarimetry
programme which targeted 573 stars brighter than V = 4\,mag to search
for magnetic fields \citep{2017sbcs.conf...86N}.  In agreement with
similar observations of 85 classical Be stars
\citep{2016ASPC..506..207W}, no large-scale field was found (Wade,
priv.\ comm.; Neiner et al., in prep.).

$\nu$\,Pup lies in the temperature range of various types of
chemically peculiar stars.  Therefore, the co-added ESPaDOnS spectrum
was inspected for traces of anomalous abundances.  No lines were found
from HgII (3984\,\AA), MnII (4137\,\AA; the lines between 3442\,{\AA}
and 3498\,{\AA} are outside the [useful] range of all three
spectrographs), CrII (4559\,\AA), SrII (4078, 4215, and 4305\,\AA),
and EuII (4205\,\AA); however, only spectral synthesis could establish
quantitative limits.  SiII\,$\lambda\lambda$4128,4131,
$\lambda\lambda$5041,5056, and $\lambda\lambda$ 6347,6371 are present
at normal strength and, in the other spectra, exhibit no variability.

The two {\it IUE} SWP high-dispersion spectra from 1988 April were
processed as described in \citet{2017ApJ...843...60W}.  Owing to the
low distance to $\nu$\,Pup, interstellar lines did not have to be
removed.  The search for an ultraviolet signature of a hot companion
used the cross-correlation technique successfully employed by
\citet{2018ApJ...853..156W}.  Tlusty \citep{2003ApJS..146..417L}
template spectra for temperatures of 27.5, 35, and 45\,kK were
employed.

\begin{table}
  \caption{Means of radial velocities of stellar Balmer lines 
    H$\gamma$, H$\delta$, H$\epsilon$, and H8, 
    of the shell absorption in H$\alpha$, and of other stellar lines 
    (He\,I\,4026, He\,I\,4143, He\,I\,4471, and Mg\,II\,4481).  
    For a description of the 
    measurement procedures see Sect.\,\ref{spectroscopy}.  
    C, E, F, and H identify the 
    spectrographs used:  CES, ESPaDOns, FEROS, and {\sc HEROS}, respectively.  
    The ESPaDOns entry is the average over 40 spectra obtained within less 
    than 1 hour.  The CES measurements are from \citet{1989A&AS...79..423B} 
    and are based on He\,I\,4471.5 and Mg\,II\,4481.2 only.  
    After $\sim$HJD\,2,451,500, emission or shell absorption 
    components of H$\alpha$ were no longer visible.  Colons (:) indicate 
    uncertain measurements.  }
\label{spectra}
\centering
\begin{tabular}{c c c c c}
\hline\hline
Spectrogr.   & HJD-2,400,000 & \multicolumn{3}{c}{Radial velocity}    \\
             &               & H$\gamma$-H8 & Shell H$\alpha$ & Other \\
             &    [d]        & \multicolumn{3}{c}{[km/s]}             \\
\hline
       C     & 46085.621     &                 &          &  37.5  \\
       C     & 46085.681     &                 &          &  38.8  \\
       C     & 46086.647     &                 &          &  34.6  \\
       C     & 46086.793     &                 &          &  35.8  \\
       H     & 49861.975     &        33.8     &   34.5   &  30.7  \\
       H     & 50160.065     &        44.6     &   34.5   &  45.9  \\
       H     & 50172.030     &        42.9     &   31.4   &  46.5  \\
       H     & 50220.962     &        38.5     &   32.8   &  35.9  \\
       H     & 50224.964     &        33.4     &   33.1   &  33.8  \\
       H     & 50227.943     &                 &   32.7   &        \\
       H     & 50228.443     &        34.5     &          &  31.7  \\
       H     & 50230.957     &        34.2     &   33.2   &  29.4  \\
       H     & 50236.952     &        33.5     &   32.1   &  29.5  \\
       H     & 50238.953     &                 &   32.8   &        \\
       H     & 50456.077     &        40.1     &   33.7   &  26.2  \\
       H     & 50467.193     &                 &   34.7   &        \\
       H     & 50471.181     &        36.6     &   34.5   &  27.9  \\
       H     & 50477.171     &        36.1     &   32.8   &  28.2  \\
       H     & 50516.011     &        34.9     &   33.9   &  24.0  \\
       H     & 50531.059     &        36.2     &   33.4   &  29.0  \\
       H     & 50551.041     &        33.8     &   35.0   &  29.8  \\
       F     & 51183.230     &        32.9     &   31.8:  &  32.1  \\
       H     & 51299.959     &        30.6     &   36.2:  &  26.3  \\
       H     & 51311.004     &        29.1     &   35.4:  &  20.2  \\
       H     & 51328.976     &        31.4     &   37.6:  &  19.6  \\
       H     & 51341.955     &        35.5     &   37.3:  &        \\
       H     & 51352.951     &        32.4     &          &        \\
       F     & 51382.446     &        29.3     &   32.8   &  32.6  \\
       F     & 51384.440     &        31.4     &   31.9   &  30.6  \\
       F     & 51564.289     &        34.0     &          &  30.3  \\
       F     & 53396.999     &        30.9     &          &  28.1  \\
       F     & 53397.000     &        30.2     &          &  25.4  \\
       F     & 53398.001     &        29.4     &          &  29.4  \\
       E     & 57411.391     &        33.7     &          &  31.0  \\
\hline
\end{tabular}
\end{table}

\subsection{Photometry:  overview}
\label{photoover}

The point-spread function of SMEI has a diameter of order one degree
\citep{2007SPIE.6689E..0CH}, and BRITE uses apertures of up to
25\,arcmin \citep{2017A&A...605A..26P}.  Within 30\,arcmin from
$\nu$\,Pup, SIMBAD \citep{2000A&AS..143....9W} lists 93 objects the
variability of which may contaminate measurements of $\nu$\,Pup.  The
brightest sources by far are the two B3\,V stars HIP\,31642
($V$\,=\,6.86\,mag) and HIP\,31875 ($V$\,=\,7.41\,mag) at distances of
16 and 25\,arcmin, respectively.  All other sources with known
brightness are of ninth magnitude or fainter, therefore each
contributing (mostly: much) less than 1\% to the total flux.  The
scatter of the Hipparcos magnitudes of HIP\,31642 (HIP\,31875) is 7.7
(9.8) mmag.  Fitted sine curves with frequency f2\,=\,0.655802\,c/d
reported below have semi-amplitudes of 2.2 and 1.3\,mmag for
HIP\,31642 and HIP\,31875, respectively.  Therefore, any corruption by
these two stars of the variabilities observed in $\nu$\,Pup is
negligible.

The BRITE and SMEI observations are highly complementary to each
other.  The former have a much higher fidelity per data point and
reveal details of the light curve while the SMEI data extend over a
much longer continuous time span and are better suited to study
long-term variations and resolve closely spaced features in the power
spectrum as has been shown in a number of independent analyses
\citep[e.g.,][]{2011MNRAS.411..162G, 2014MNRAS.445.2878H,
  2018A&A...610A..70B}.  The noise and the time span of the Hipparcos
data are intermediate between those of SMEI and BRITE whereas the
number of Hipparcos data points is one to two orders of magnitude
smaller. Accordingly, the Hipparcos photometry can be used to trace
back, over more than a decade, major findings made with the other
satellites.

Simple inspection of the non-phase-folded BRITE light curve
(Fig.\,\ref{LCmulti}) seems to suggest that it is single-periodic, not
sinusoidal, and, on timescales of months or years, strongly variable
in amplitude.  There are indications that consecutive light cycles do
not repeat perfectly (Fig.\,\ref{BLC}).  Compared to other Be stars,
the BRITE power spectrum of \object{$\nu$\,Pup} (Fig.\,\ref{PS}) also
looks unusually simple with only one very dominating feature near
0.656\,c/d.  However, the sensitivity of {\it MOST}, {\it Kepler}, and
CoRoT to variabilities of B-type stars (cf.\ Introduction) was at least
two orders of magnitude better than that of BRITE, which did not reach
sub-mmag levels for $\nu$\,Pup.  Moreover, the simplicity may be just
apparent because time-series analyses of different segments of the
BRITE photometry return very substantially different frequencies.
According to Fig.\,\ref{BLC}, there is no appreciable phase difference
between the variabilities in the blue and the red BRITE passbands so
that combining all BRITE data is not expected to introduce artifacts
into the power spectra.

\begin{figure}
\includegraphics[width=9.9cm,angle=-90]{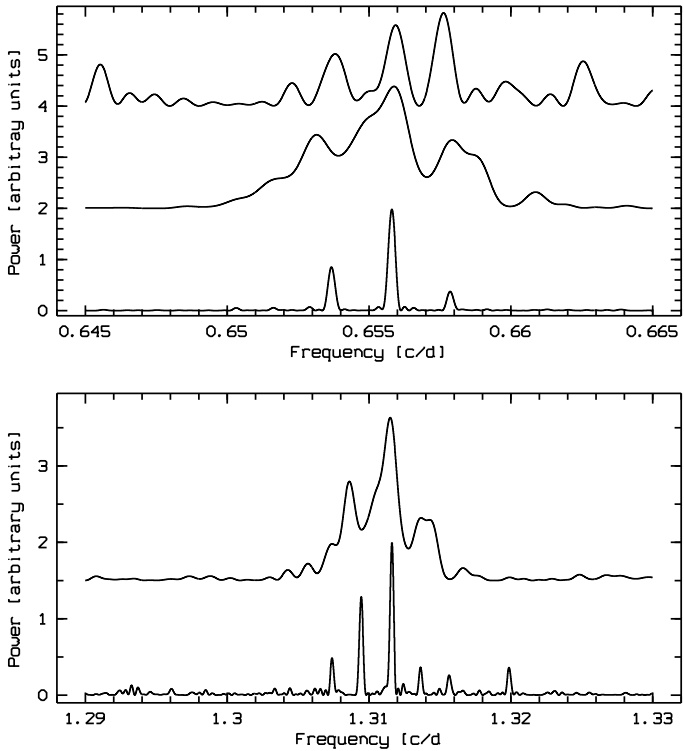}
\caption{Upper panel: Power spectra of Hipparcos (top), BRITE
  (middle) and SMEI (bottom, with frequencies c1 to c6
  [Table\,\ref{frqs}] appearing between 1.3075 and 1.32\,c/d)
  observations.  The three peaks in the SMEI power spectrum correspond
  with frequencies f1 to f3 (Table\,\ref{frqs}).  Lower panel: SMEI
  (bottom) and BRITE (top) power spectra of the frequency group near
  1.31\,c/d.  The frequency scale is stretched by a factor of 2
  w.r.t.\ the upper panel so that base frequencies and harmonics line
  up.  The power scales are arbitrary and cannot be compared between
  the satellites.  The SMEI power scales for the two frequency groups
  differ by about a factor of 20.}
\label{PSzoom} 
\end{figure}


The SMEI power spectrum reveals the same overall structure as that
obtained with BRITE (Fig.\,\ref{PS}).  Apart from the 0.656-c/d
variability, significant other features in the
SMEI power spectrum are again only the first harmonic, 1\,c/d, and
1-c/d aliases.  Higher harmonics were not detected.  The
primary feature near 0.656\,c/d and its harmonic near 1.31\,c/d are
well resolved (Fig.\,\ref{PSzoom}), explaining the variable frequency
in shorter BRITE data segments.  The power spectrum can be described
as consisting of two frequency groups that are extremely narrow.  The
main frequency is f2\,=\,0.655802\,c/d.  All frequencies are listed in
Table\,\ref{frqs}; they are analyzed in Sects.\,\ref{coarse} and
\ref{simulations}.

The observations with both BRITE and Hipparcos alone do not adequately
resolve the two frequency groups (Fig.\,\ref{PSzoom}).  Simulations
confirm that the BRITE observations sample the $\sim$480-d beat period
(Fig.\,\ref{VarAmpSMEI}) very unfortunately.  However, when the BRITE
data are added to the SMEI dataset (see below), all peaks in the power
spectrum (frequencies f1 to f3 and c1 to c6 in Table\,\ref{frqs})
become narrower and higher, demonstrating that both datasets have
these frequencies in common.

The results of single-frequency fits to the BRITE observations are
included in Table\,\ref{frqs}.  At 0.65587(6)\,c/d and with an
amplitude of 12(2)\,mmag, f2 is clearly detected also with Hipparcos, and
there may be peaks corresponding to f1 and f3.  Accordingly, at
least f2 existed about a decade before the beginning of the SMEI
observations.  Without the BRITE and SMEI observations, f2 could not
have been confidently identified with Hipparcos alone because it is
well above any possible Nyquist frequency.  This probably explains the
suggestion by \citet{2002MNRAS.331...45K} of a 0.15292-c/d frequency,
which is relatively close to 0.656\,c/d$ - 0.5$\,c/d.  This frequency
folds the Hipparcos data to a reasonable-looking light curve (with
some suspicious fine structure); however, the same is not true for
both BRITE and SMEI.

Time-series analyses were also performed for the data combinations
SMEI+BRITE and SMEI+BRITE+Hipparcos.  As expected, the results hardly
differ between these two datasets, and the addition of BRITE and/or
Hipparcos data does not qualitatively alter the power spectra shown in
Fig.\,\ref{PSzoom}.  With BRITE added, the full width at half maximum
of the amplitude profile of frequency f2 (Table\,\ref{frqs}) decreases
from 0.00048\,c/d to 0.00020\,c/d.  In synthetic data with constant
frequencies f1 to f3, the change is only from 0.00046\,c/d to
0.00034\,c/d, in agreement with the full time span of 5200\,d.
The value of 0.00020\,c/d concerns a narrow spike located above a much
broader pedestal (see also Fig.\,\ref{photoDoppler}).  In any event,
there is no quantifiable indication that the width of f2 is broadened
by any unidentified process or an unresolved component.

Because of differences in spectral throughput, noise, and data
processing, a direct comparison of the amplitudes measured with BRITE,
SMEI, and Hipparcos is not advisable.  For an interpretation of
blue-to-red amplitude ratios, a calibration is missing.

\subsection{Photometry: frequency groups}
\subsubsection{Description}
\label{coarse}

In the SMEI data, the main variability at f2\,=\,0.655802(3)\,c/d
(numbers in parentheses are 1-$\sigma$ uncertainties of the last digit
of measured values) has a mean amplitude of 6.6$\pm$1.4\,mmag.  Two
other peaks appear at f1\,=\,0.653676(4) and f3\,=\,0.657861(6)\,c/d
(Fig.\,\ref{PSzoom}) with amplitudes of 4.3$\pm$1.4 and
2.9$\pm$1.5\,mmag, respectively.  The differences in frequency from
the central peak are $-$0.00213\,c/d and $+$0.00206\,c/d,
respectively.  That is, perfect symmetry is not strictly excluded.

The bottom panel of Fig.\,\ref{PSzoom} zooms in on the power spectrum
in the region of the second frequency group around 1.31\,c/d.  For
SMEI, peaks at c1\,=\,1.30740(3), c3\,=\,1.31160, and
c5\,=\,1.31565\,c/d correspond to within the errors to the harmonics
of f1, f2, and f3, respectively.  Two additional peaks at
c2\,=\,1.30944(2) and c4\,=\,1.31363(3) are about equidistantly
inserted between the other three.  Although c2 at 1.30945\,c/d is the
second strongest feature of the five, it is without any equivalent at
half this frequency.  That is, it cannot be a harmonic under the
assumption of constant frequencies.  The same holds for the second of
the intermediate frequencies (c4).  The 1.31-c/d group includes a sixth
frequency somewhat offset from the others at c6\,=\,1.31987(3)\,c/d.

\begin{table*}
  \caption{The values of the frequencies visible in Fig.\,\ref{PSzoom} 
    and derived from single-frequency sine fits.  
    f1 to f3 are the base frequencies, c1 to c5 the harmonic/sum frequencies.  
    The latter are also calculated from f1 to f3 with the shown  
    formulae.  Column $\Delta$f contains the differences between the 
    observed frequencies in the lines directly above and below.  
    The amplitudes, especially of the combination frequencies, 
    have large uncertainties.  }
\label{frqs}
\centering
\begin{tabular}{c c c c c c c | c c c}
\hline\hline
      & \multicolumn{6}{c}{SMEI}                                                          & \multicolumn{3}{c}{BRITE} \\
ID    & SMEI        &  $\Delta$f & Calculation    &  Calculated & Difference & SMEI       & BRITE        & $\Delta$f & BRITE      \\ 
      & frequency   &            & scheme         &  frequency  & O-C        & amplitude  & frequency    &           & amplitude  \\
      &  [c/d]      &    [c/d]   &                &     [c/d]   &   [c/d]    &   [mmag]   &   [c/d]      &   [c/d]   &   [mmag]   \\
\hline
f1    & 0.653676(4) &            & N/A            &  N/A        &  N/A       & 4.3        & 0.65314(2)   &    N/A    &    7.6      \\
      &             &  0.00213   &                &             &            &            &              &           &             \\
f2    & 0.655802(3) &            & N/A            &  N/A        &  N/A       & 6.6        & 0.65880(1)   &    N/A    &    9.8      \\
      &             &  0.00206   &                &             &            &            &              &           &             \\
f3    & 0.657861(6) &            & N/A            &  N/A        &  N/A       & 2.9        & 0.65793(2)   &    N/A    &    7.2      \\
c1    & 1.30740(3)  &            & 2*f1           &  1.30735    & $+0.00005$ & 0.7        & 1.3074(1)    &           &    1.2      \\
      &             &  0.00204   &                &             &            &            &              &   0.0012  &             \\
c2    & 1.30944(2)  &            & f1+f2          &  1.30948    & $-0.00004$ & 1.2        & 1.30861(6)   &           &    2.0      \\
      &             &  0.00216   &                &             &            &            &              &   0.00278 &             \\
c3    & 1.31160(1)  &            & 2*f2           &  1.31160    & $+0.00000$ & 1.4        & 1.31149(5)   &           &    2.6      \\
      &             &            & f1+f3          &  1.31154    & $+0.00006$ &            &              &           &             \\
      &             &  0.00203   &                &             &            &            &              &   0.00220 &             \\
c4    & 1.31363(3)  &            & f2+f3          &  1.31366    & $-0.00003$ & 0.6        & 1.31369(8)   &           &    2.0      \\
      &             &  0.00202   &                &             &            &            &              &           &             \\
c5    & 1.31565(4)  &            & 2*f3           &  1.31572    & $-0.00007$ & 0.5        &              &           &             \\
c6    & 1.31987(3)  &            & 2*f2+4*(f3-f2) &  1.31984    & $+0.00003$ & 0.6        &              &           &             \\
\hline
\end{tabular}
\end{table*}

\begin{figure}
\includegraphics[width=14.2cm,angle=-90]{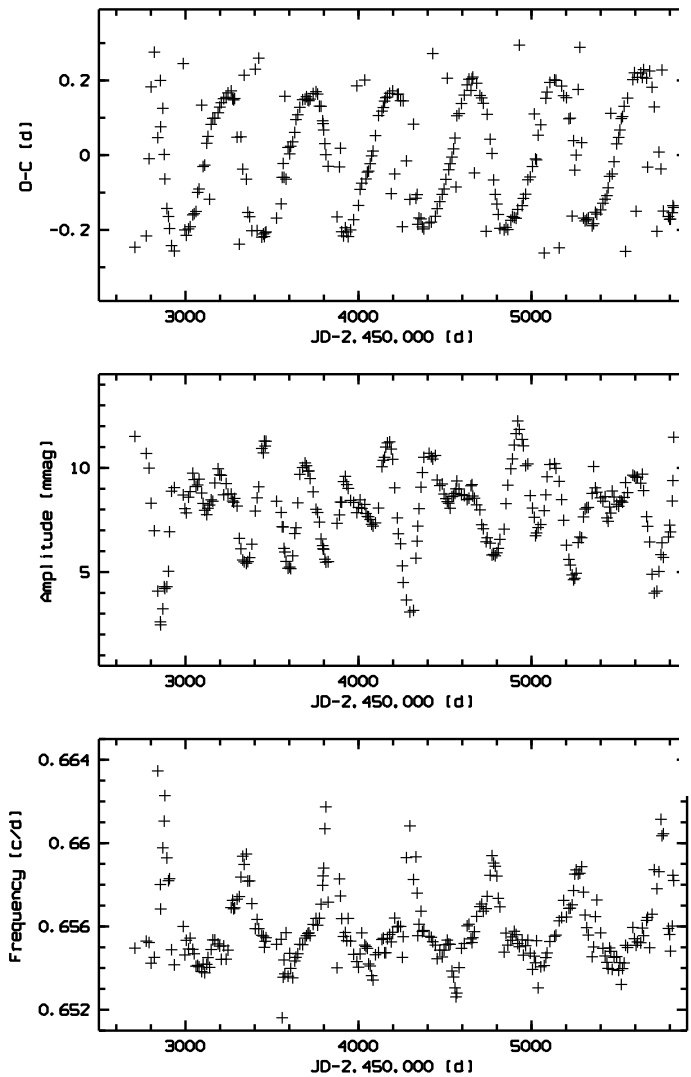}
\caption{Bottom: Frequencies of single sine functions fitted to the
  SMEI data of $\nu$\,Pup over time intervals of 75\,d stepped by
  10\,d.  Middle: Ditto except for amplitudes.  Top: O-C curve for a
  mean frequency of 0.655802\,c/d.  In all three panels, outliers have
  been clipped.}
\label{VarAmpSMEI} 
\end{figure}

\begin{figure}
\includegraphics[width=9.7cm,angle=-90]{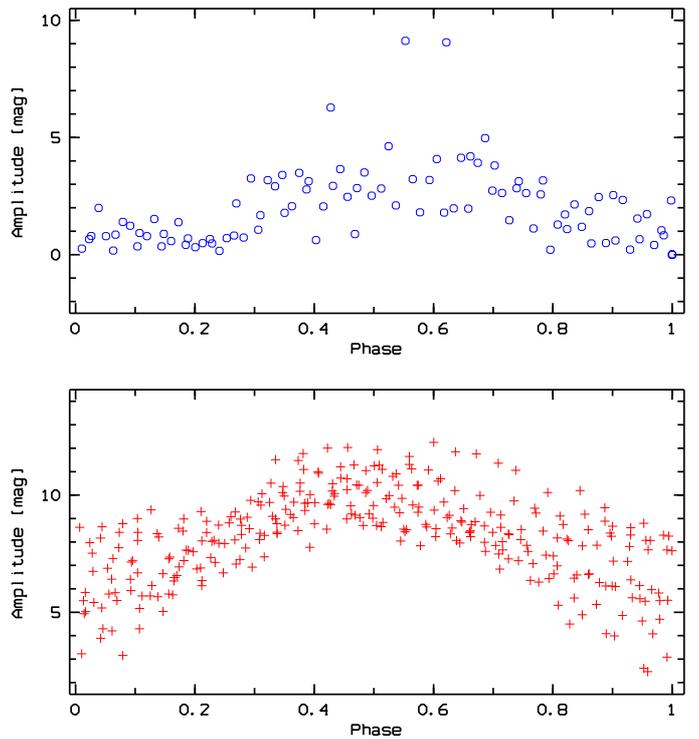}
\caption{Bottom: The amplitudes from Fig.\,\ref{VarAmpSMEI} phased
  with a frequency of 0.004186\,c/d.  Top: Fitted amplitudes of the
  bottom panel phased with a frequency of 0.00130\,c/d.}
\label{ampamp} 
\end{figure}

\begin{figure}
\includegraphics[width=14.2cm,angle=-90]{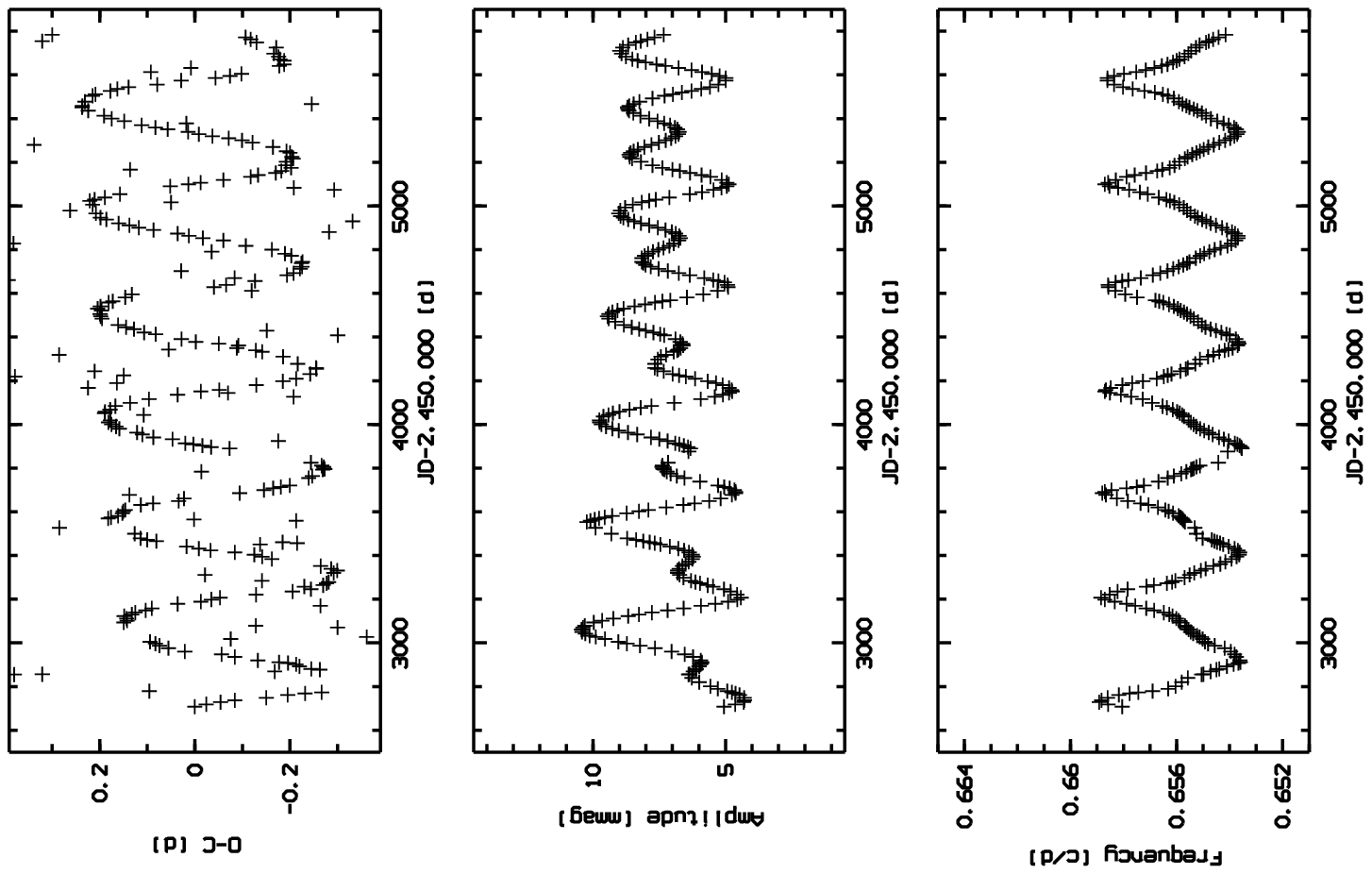}
\caption{Same as Fig.\,\ref{VarAmpSMEI} except for synthetic data as 
described in Sect.\,\ref{simulations}.}
\label{VarAmpSynth} 
\end{figure}

Table\,\ref{frqs} discloses that all SMEI frequencies in the group
around 1.31\,c/d can be derived with high fidelity from those in the
group near 0.656\,c/d.  The frequencies c2 and c4 between the three
harmonics c1, c3, and c5 are equal to the sum frequencies f1+f2 and
f2+f3, respectively, to within the errors; they do not have
counterparts in the 0.656-c/d group.  Finally, the somewhat more
distant c6 can be expressed as 2$\times$f2+4$\times$(f3$-$f2).  Given
the numerous relations between all frequencies, this latter choice is
necessarily not unique.

The described frequency scheme also suggests that - at the given
photometric detection level - the groups of f1 to f3 and c1 to c5 are
complete because the scheme and the frequencies in one group cannot be
used to predict frequencies in the respective other group that are not
observed.  However, the presence of c6 suggests that a more complex
extension to, or replacement of, the core scheme exists.  In any
event, additional variabilities will probably hide below the detection
threshold of the BRITE and SMEI data.  The variable shape of
individual cycles of the light curve (Figs.\,\ref{LCmulti} and
\ref{BLC}) is presumably caused by the multiperiodicity.  The
persistence of the harmonic frequencies implies persistent asymmetries
in agreement with Fig.\,\ref{LCmulti}.

A structure with two symmetric side lobes can arise in the power
spectrum if a frequency and/or amplitude is modulated with the
frequency difference between the central peak and the other two.  If
f1 to f3 are exactly equidistant, as marginally permitted but not
required by the time-series analysis, it needs to be determined
whether the variability is due to a single but periodically modulated
frequency or due to three separate frequencies.  If only frequencies
are considered, the problem is degenerate.  An apparent advantage of
the single-modulated-frequency hypothesis is that it easily explains
c1 to c5 if c3 is the harmonic of f2.  Obviously, three separate
frequencies cannot achieve this.  c6 remains unexplained in both
cases.

As a first step towards resolving this ambiguity, single-frequency
sine fits to the SMEI data were performed in 75-d windows stepped by
10\,d.  The window of 75\,d was chosen because it is about the
shortest time interval over which meaningful fits of the 0.656\,c/d
variability can be obtained.  It falls far short of resolving the
0.0021\,c/d spacing.  Fig.\,\ref{VarAmpSMEI} presents the results.  The
bottom panel suggests that the frequency is modulated with a
quasiperiod near 480\,d (0.0021\,c/d).  From the middle panel, one
would infer that the amplitude varies about twice as fast as the
frequency does.  This double frequency is confirmed by a time-series
analysis (Fig.\,\ref{ampamp}).  The amplitude of this amplitude
modulation also varies cyclicly with 0.00013\,c/d
(Fig.\,\ref{ampamp}).  For a variable frequency, an O-C curve can be
constructed which illustrates the difference between the observed
phase (O; for instance measured as the times of maximal brightness)
and that calculated (C) for the mean frequency.  The O-C curve for
f2\,=\,0.655802\,c/d is depicted in the top panel, which seems to
imply that there is a periodic $\pm0.2$-d phase wobble.

Because frequencies f1 to f3 alone are not sufficient to discriminate 
between one modulated and three separate frequencies, the usage for the
above analysis of a single modulated frequency could be seen as
introducing a bias.  However, this choice is without pre\-judice to this
issue, and the structures in Fig.\,\ref{VarAmpSMEI} will be of decisive 
diagnostic value in the analysis below (Sect.\,\ref{simulations}).

\subsubsection{Simulations}
\label{simulations}

Figure\,\ref{VarAmpSMEI} makes it clear that a single modulated
frequency would require a complex physical model to explain the large
and conspicuous variations in amplitude and frequency.  Moreover, the
frequencies of these two quantities differ by a factor of $\sim$2
(Figs.\,\ref{VarAmpSMEI} and \ref{ampamp}).  Conjectures about such a
model are deferred until simulations have investigated whether the
structures in Fig.\,\ref{VarAmpSMEI} are compatible with the
alternative hypothesis of three separate frequencies.  Because the
observed frequencies are not unresolved blends
(Sect.\,\ref{photoover}) and the sets of f1 to f3 and c1 to c5 appear
complete (Sect.\,\ref{coarse}), simulated data should achieve
acceptable approximations of the reality.

\begin{figure}
\includegraphics[width=12.7cm,angle=-90]{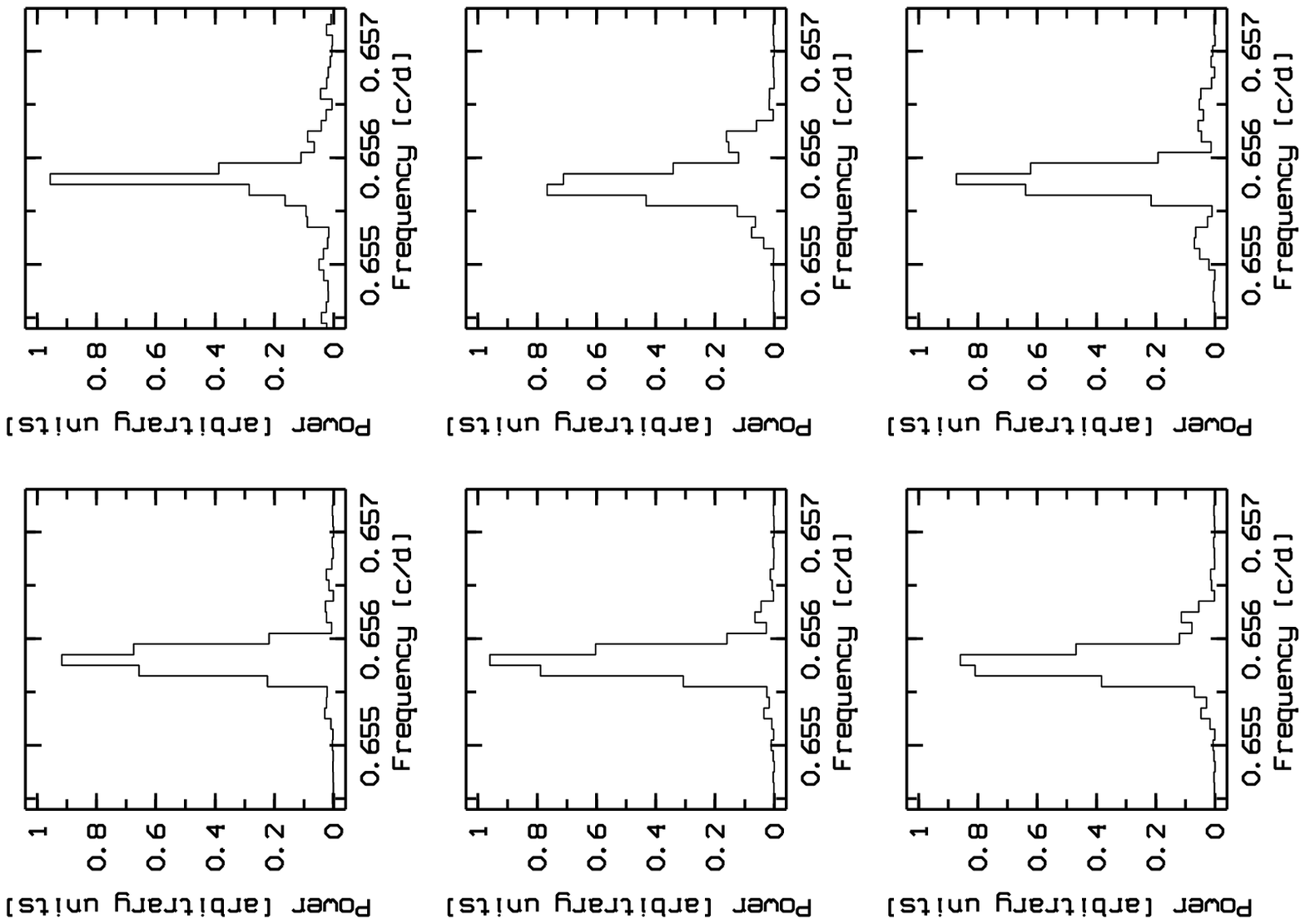}
\caption{Top left: Power spectrum of SMEI observations between
  frequencies f1 and f3 (Table\,\ref{frqs}).  Top right: Ditto, except
  for combined SMEI and BRITE data.  The other four panels show power
  spectra of a single sine function with frequency f2 evaluated at
  SMEI observing times.  This variation is modulated with different
  amplitudes (A) and frequencies (F).  0.0003125\,c/d is about the
  inverse of the time span of the SMEI observations.  Middle left:
  A\,=\,10\,km/s, F\,=\,0.0003125\,c/d; there is almost no difference
  w.r.t. the SMEI observations.  Middle right: A\,=\,30\,km/s,
  F\,=\,0.0003125\,c/d; the strength of the side lobes increases with
  the amplitude.  Bottom left: A\,=\,20\,km/s, F\,=\,0.0003125\,c/d;
  bottom right: A\,=\,20\,km/s, F\,=\,0.000625\,c/d. The separation
  of main peak and side lobes increases with modulation
  frequency.}
\label{photoDoppler} 
\end{figure}

\begin{figure}
\includegraphics[width=4.1cm,angle=-90]{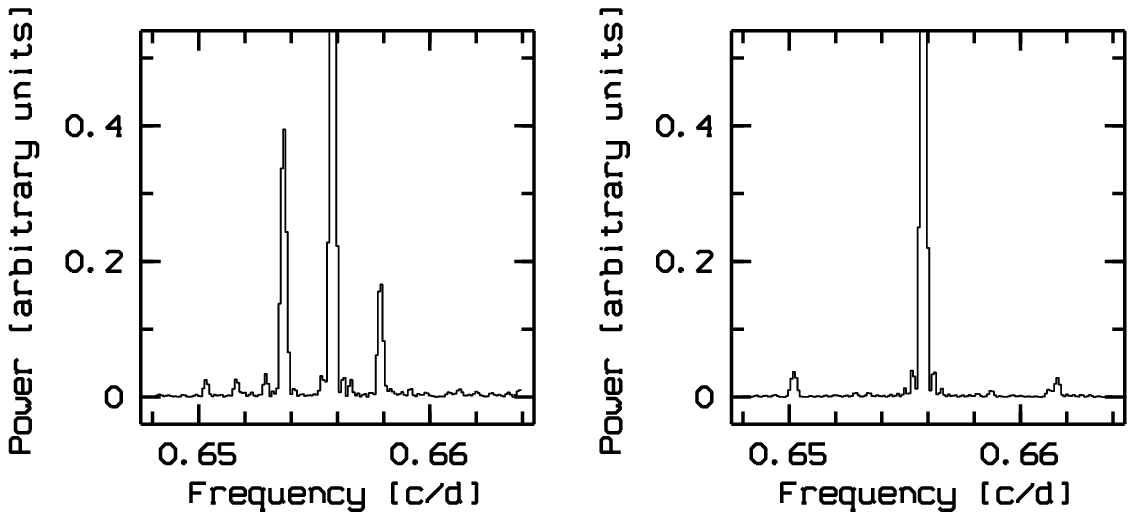}
\caption{Left panel: Same as top left panel of
  Fig.\,\ref{photoDoppler} except for compressed abscissa.  Right
  panel: Same as lower four panels of Fig.\,\ref{photoDoppler} except
  for compressed abscissa and A\,=\,12\,km/s and F\,=\,0.0057\,c/d,
  corresponding to a 1-M$_{\odot}$-star orbiting $\nu$\,Pup at 1\,au.
  In the left panel, f1 to f3 (Table\,\ref{frqs}) are visible whereas
  the synthetic data in the right panel only include f2 and its side
  lobes due to the assumed orbital motion.}
\label{photoDmodel} 
\end{figure}


A model with three constant frequencies (f1 to f3), i.e.\, omitting
the harmonics, was built from sine functions evaluated at the actual
SMEI observing times.  Amplitudes and phases were derived from
single-frequency sine fits to the observations.  The data were
analyzed in exactly the same way as the real SMEI observations
(Sect.\,\ref{coarse}).  They are plotted in Fig.\,\ref{VarAmpSynth},
which is not a fit but a proof of concept and as such demonstrates
beyond any doubt that three separate frequencies can reproduce
Fig.\,\ref{VarAmpSMEI} in much detail.

In conclusion, the simulated data strongly favor the
three-separate-frequencies option.  An important and very firm outcome
of the simulations is that the frequency doubling in the amplitude
curve disappears when f1 to f3 are truly equidistant.  This is also
analytically clear because, then, only one beat frequency exists.
Therefore, f1 to f3 are not equidistant, which eliminates the need for
the single-modulated-frequency conjecture altogether.

\subsection{Mass loss}
\label{massloss}

While BRITE can be effective at detecting photometric signatures of
outbursts \citep{2017arXiv170808413B, 2018A&A...610A..70B}, the
extensive clipping applied to the SMEI observations of $\nu$\,Pup may
well have led to the removal also of genuinely stellar events
$\lesssim$5\,mmag.  For the BRITE data of $\nu$\,Pup, the detection
threshold is lower but not by much because, for the stitching together
of the numerous short data strings, it was assumed that between
consecutive datasets there were no jumps intrinsic to the star.  

By coincidence, the BHr2018 observations in the two bottom panels of
Fig.\,\ref{LCmulti} not only have the lowest instrumental scatter but
were also obtained when the total photometric amplitude of $\nu$\,Pup
was lowest.  These data should have the highest sensitivity to
outbursts.  Three peaks, consisting of four data points each, stand
out above the ambient light curve by 5\,mmag.  They are the only
candidate brightenings possibly related to mass loss that could be
found.  As the BHr2017 observations in Fig.\,\ref{BLC} confirm, a
5-mmag offset of four consecutive data points each is significant,
and the brightenings occurred well within an uninterrupted series of
observations.  However, non-Gaussian errors do occur, and it would
require simultaneous observations with two instruments to ascertain
(or not) the reality of such events.  Owing to the beating with much
larger amplitude, even two BRITE satellites would probably find it
difficult to identify events of similar magnitude but ten times longer
duration as seen in the B5e star KIC\,11971405
\citep{2015MNRAS.450.3015K, 2016A&A...593A.106R}.

\subsection{Binarity}
\label{Doppler}

A late-type, i.e.\ low-mass, Be star such as $\nu$ Pup might offer an
opportunity to derive useful constraints on the presence of a
companion star.  Pulsation-induced variations and circumstellar
components of line profiles are a lesser problem than in early-type Be
stars.  However, strong rotational line broadening and the paucity of
suitable spectral lines pose other challenges, as also seen in
$\nu$\,Pup.  Photometric Doppler variations can, therefore, be a
useful alternative to conventional spectroscopy, especially since
accurate measurements of multi-frequency variations necessarily
require excellent phase coverage.

An elaborate description of the power of photometric Doppler shifts
for the analysis of binaries with pulsating component(s) is available
from \cite{2015MNRAS.450.3999S} and references therein.  Here, only
the most basic application is used.  Figure\,\ref{photoDoppler}
derives, in the low-frequency regime, an upper limit of $\sim$20\,km/s
on the amplitude of sinusoidal radial-velocity variations and
illustrates that frequencies down to about 0.0006\,c/d (periods up to
$\sim$4.5 years) can be excluded.  The location of f1 and f3 at a
separation from f2 of about $\pm$0.0021\,c/d (see left panel of
Fig.\,\ref{photoDmodel}) leads to a blind spot of the method around
that frequency.  The right panel of Fig.\,\ref{photoDmodel} presents
the simulated power spectrum around f2 for a 1-M$_{\odot}$ star
orbiting $\nu$\,Pup at a distance of 1 au (period nearly half a year;
frequency $\sim$ 0.0057\,c/d).  Because of the extreme contrast in the
observations between the peak at 0.655802-c/d and all other features
(see left panel of Fig.\,\ref{photoDmodel}), it would be possible to
identify the two weak spikes at 0.6558$\pm$0.0057\,c/d.

\begin{figure}
\includegraphics[width=7.4cm,angle=90]{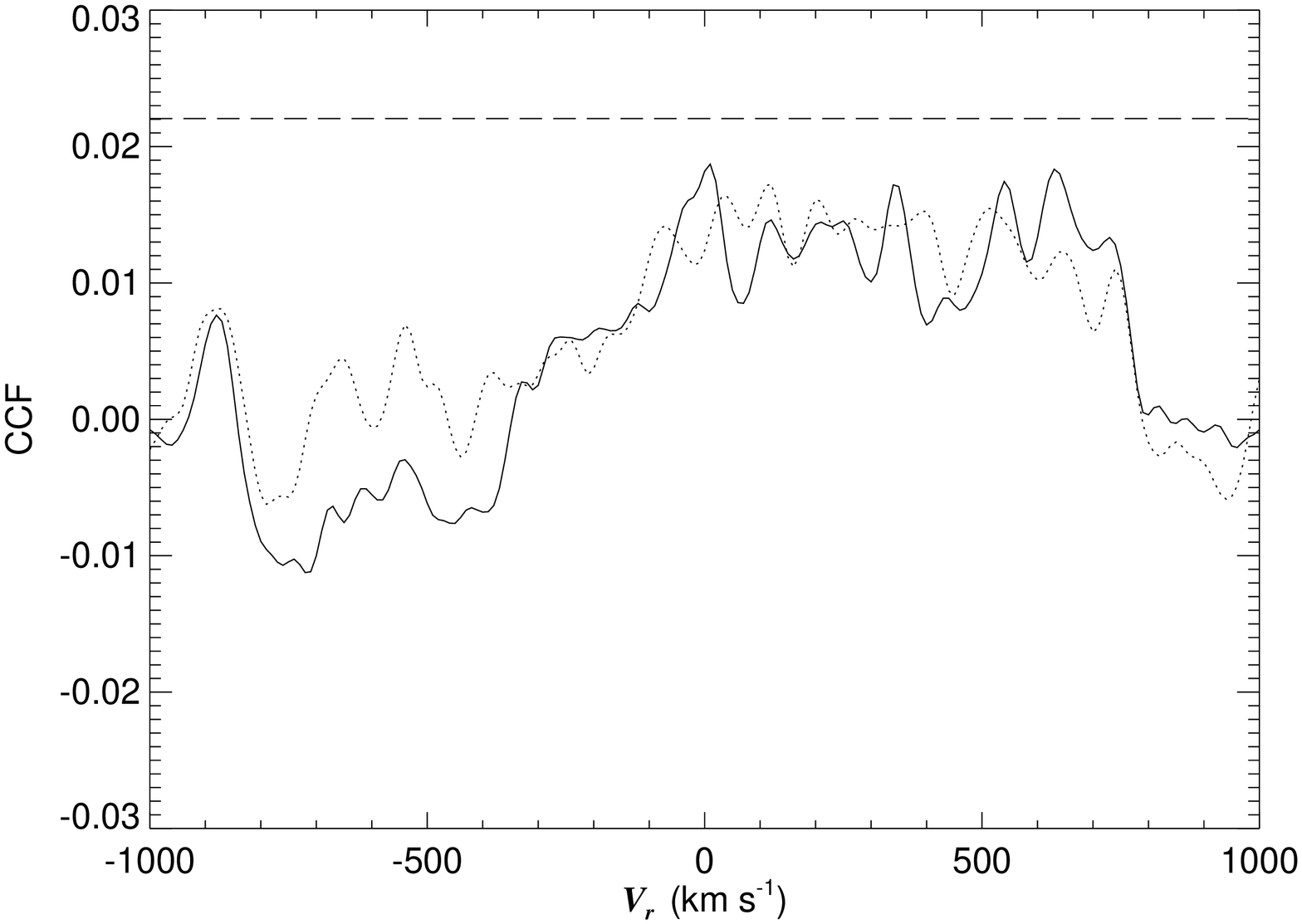}
\caption{Cross-correlation functions between the two {\it IUE} observations of
  $\nu$\,Pup and the 45\,kK model template as described in
  Sect.\,\ref{Doppler}.  Because of different degrees of similarity to
  the spectrum of the Be primary, the cross-correlation functions for
  the 45\,kK template are narrower than those with the 35\,kK and
  27.5\,kK templates. The horizontal line indicates a signal-to-noise
  level of 3.0 above which a companion with temperature $\sim$45\,kK
  would be considered significant.}
\label{IUE} 
\end{figure}

In the {\it IUE} spectra, evidence of a hot companion was not found.
From cross-correlation of the observations with the hottest model
template spectrum (45\,kK) only an upper limit of $\sim$2\% of the
flux of the Be primary could be derived for the secondary.  The
signal-to-noise ratio of 2.49 is below the significance threshold of
3.0 established by \citet{2018ApJ...853..156W}.  The low temperature
of the primary also enabled the usage of templates corresponding to
temperatures of 35 and 27.5\,kK.  The resulting cross-correlation
functions are broader than that with the 45\,kK template (Fig.\,\ref{IUE})
because of stronger similarities with the B-type primary. In the
classification scheme of \citet{2018ApJ...853..156W}, $\nu$\,Pup would
be assigned code 'P' (cross-correlation signal from primary star).

Finally, the good agreement between the Hipparcos and the Gaia DR2
parallaxes also indicates that, on timescales of few years, there is
no strong orbital reflex motion caused by an unseen companion.

\section{Discussion}
\label{discussion}

$\nu$\,Pup is a normal B7-B8\,IIIe shell star.  A magnetic field was
not detected.  High-quality spectra exhibited no abundance anomalies
or other pecularities.  The upper limit on periodic sinusoidal RV
variations is 10\,km/s (Sect.\,\ref{spectroscopy}).  For amplitudes in
that range, periods up to a few years appear ruled out
(Sect.\,\ref{Doppler}).  Weak emission and shell absorption components
in H$\alpha$ \citep{1999A&A...348..831R} were observed between 1995
and 1999; they may have marked the end of the disk dissipation period
after stronger emission a decade before.

The detection with photographic plates of shell absorption in
H$\gamma$ \citep{1909ApJ....29..232C} implies that either 80-90
years earlier the circumstellar disk was much more developed or in the
1990's the peak of the emission phase was, in fact, missed.  The complete
absence of H$\alpha$ line emission in the years 2000, 2005, and 2016
(see Table\,\ref{Halpha}) may be a weak indication that a good part of
the SMEI (2003-2011) and BRITE (2015, 2016/17, and 2017/18)
observations perhaps took place while $\nu$\,Pup did not maintain a
disk of much significance.  For the Hipparcos data, no parallel
spectroscopic record exists; however, in the 1989-1993 time interval
the disk may have been dissipating (Sect.\,\ref{spectroscopy}).
Accordingly, the persistent, regular variability seen with SMEI or
BRITE should be of primarily stellar -- instead of circumstellar --
nature.

The tight and homogeneous frequency pattern evidences that the
entire variability is intrinsic to one single star.  There is no
contamination by a companion or field star.  Frequencies f1 to f3 and
c1 to c5 seem to be complete sets in that no frequencies following
the construction scheme of Table\,\ref{frqs} are missing
(Sect.\,\ref{coarse}).  The frequencies found are not compromised by
unresolved blends (Sect.\,\ref{photoover}).

Simulations (Sect.\,\ref{simulations}) have demonstrated that the
0.656-c/d variability consists of three separate frequencies (f1 to
f3).  There is also no physical motivation to consider frequency
modulation with 0.0021\,c/d: Binary models are rejected in
Sects.\,\ref{Doppler} and \ref{binarityD}, and 0.0021\,d cannot
possibly be the light travel time in a binary because the implied mass
would be superstellar.  Oblique rotators are ruled out because a
splitting by 0.0021\,c/d would imply a rotation rate that is two
orders of magnitude below the observed line width.  A rapidly rotating
magnetic oblique pulsator would require a very strong magnetic field
that is not observed.  A complex differential-rotation scheme can be
construed but the fine tuning introduces more challenges than it can
eliminate.  Intrinsic frequency variations are known from other stars
and mostly thought to result from pulsation-mode interactions.  The
most notable example is the Blazhko effect, e.g., in RR\,Lyr
\citep{2017MNRAS.466.2602P} and $\delta$\,Cephei
\citep{2016pas..conf...22S} stars.  However, it seems to involve one
or more radial modes, which are not known from Be stars
(Sect.\,\ref{656}).

In conclusion, not only does the three-separate-frequencies model
reproduce the observations well but there is also no physical need to
add extrinsic or intrinsic long-term variability.  Therefore, the
interpretation is adopted that the 0.656-c/d variability of $\nu$\,Pup
consists of three separate variations.  Without regard to this
splitting, possible identifications of this variability are briefly
discussed in the next subsection.

\subsection{Possible causes of the 0.656-c/d variability}
\label{656}

\noindent
{\bf Radial pulsation:} 
There is apparently no case of reported radial pulsation in Be as well
as SPB stars.  This is in agreement with model calculations which
consistently find radial modes strongly damped at the temperature of
$\nu$\,Pup \citep[e.g.,][]{1999AcA....49..119P}.  For $\nu$\,Pup, a
drastic mismatch exists w.r.t.\ the mean pulsation constant
Q\,=\,0.033\,d for the radial modes of $\beta$ Cep (and similarly many
other) stars \citep{2005ApJS..158..193S}.

In conclusion, the 0.656-c/d frequency of $\nu$\,Pup is not due 
to radial pulsation.  

\noindent 
{\bf Circumstellar {\v S}tefl frequency:} 
Some Be stars exhibit a so-called {\v S}tefl frequency
\citep{1998ASPC..135..348S, 2016A&A...588A..56B}.  In spectroscopy,
{\v S}tefl frequencies are most prominent in lines formed in the
photosphere-to-disk transition region.  Their amplitude seems to trace
the star-to-disk mass transfer rate as is inferred from the
nondetection of {\v S}tefl frequencies at times when the H$\alpha$
line emission is weak or fading.  However, the yearslong absence in
$\nu$ Pup of H$\alpha$ line emission and especially the concomitant
disappearance and subsequent absence for perhaps 15$+$ years also of
shell absorption in this equator-on star argues against the
interpretation as a {\v S}tefl frequency of the 0.656-c/d variability.
In conclusion, the hypothesis of a circumstellar {\v S}tefl frequency
for the 0.656-c/d variability lacks merit.

\noindent
{\bf Rotational modulation:} 
The initial confusion of nonradial pulsation with magnetic rotational
modulation \citep[][and references therein]{2013ASSP...31..247B} has
long been resolved \citep{2013ASSP...31..253R}.  Nevertheless, the
possibility of rotational modulation cannot be fundamentally excluded.
Often in Be stars, plausible rotation frequencies are embedded in 
frequency groups which makes the identification of the former somewhat
arbitrary \citep[e.g.,][]{2009A&A...506..125D}.  Beyond such
coincidences, rotational modulation of specific properties (e.g., line
strengths of specific elements) has not been reported.  In $\nu$ Pup,
the observed frequency of 0.656\,c/d is straddled by the model
rotation rate of 0.5\,c/d (Table\,\ref{stellarPar}) and the rotation
rate 0.77\,c/d derived from the observed rotational velocity of
246\,km/s \citep{2002A&A...381..105R} and the also calculated
equatorial radius of 6.85\,R$_\odot$ (which is supported by the
observed parallax).

In conclusion, rotational modulation is well possible but not
positively demonstrated.  The lack of spectroscopic fiducial marks
(Sect.\,\ref{spectroscopy}) carried around the star is an obstacle.

\noindent{\bf Ellipsoidal variability in a binary:} In such systems,
the photometric frequency is usually twice the orbital frequency so
that in $\nu$\,Pup the orbital frequency would be 0.33\,c/d.  If the
putative companion is a white dwarf with mass 0.5\,M$_\odot$ in a
circular orbit, the orbital velocity of the Be primary would be about
30\,km/s.  Although noisy, the radial-velocity curve in
Fig.\,\ref{BalmerRV} and the photometric Doppler simulations
(Sect.\,\ref{Doppler}) do not harbor a 2\,$\times$\,30 km/s
variability. The expected semi-amplitude could be lower if
spin and orbit are misaligned.  However, the compilation by
\citet{2011ApJ...726...68A} suggests that major misalignments do not
occur in binaries of similar period.  The kink near the maximum of the
light curve (Fig.\,\ref{BLC}) could also be a secondary minimum which
could be shallower than the primary minimum if the photospheric
regions facing towards and away from the companion have different
surface brightnesses.  In that case, the orbital frequency would be
$2\times 0.33$\,c/d, much aggravating the radial-velocity problem, and
the separation of the two stars would shrink to little more than the
equatorial radii in Table\,\ref{stellarPar}.  The emission could,
then, only arise in a circumbinary disk formed by Roche-lobe overflow.

In conclusion, $\nu$\,Pup is not likely to be an ellipsoidal variable
in agreement with the general arguments presented in
Sects.\,\ref{Doppler} and \ref{binarityD} against $\nu$\,Pup being a
double star.

\subsection{Nonradial pulsation}
\label{NRP}
None of the options considered for the 0.656-c/d variability
(Sect.\,\ref{656}) has a built-in mechanism to explain the splitting
into f1 to f3, thereby further aggravating the strong tension with the
observations.  Frequency groups are widespread in Be stars (cf.\
Introduction) and are commonly attributed to multi-mode nonradial
pulsation.

Frequency groups can be difficult to delineate but typically three of
them are found.  \citet{2015MNRAS.450.3015K} suggested that the grouping
could arise from combination frequencies.  \citet{2017A&A...598A..74P}
found this tendency confirmed among the more than 1000 frequencies
diagnosed in the same data.  The preliminary picture presented by
\citet{2017arXiv170808413B} of the B1\,Ve star 25\,Ori is similar.
There is no explanation yet of the formation, prominence, and role of
combination frequencies in Be stars as pictured below.  The most
elementary empirical description that can be given of frequency groups
is that one group (g1) consists of the NRP frequencies proper and
groups g0 and g2 are built from differences and sums/harmonics,
respectively, of frequencies in g1.  However, in the presence of noise
and finite frequency resolution, a firm and satisfactorily complete
classification of a rich frequency spectrum is not straightforward.

The small number of frequencies detected in $\nu$\,Pup facilitates
this task a lot, and Table\,\ref{frqs} provides the most complete
characterization to date of frequency groups in Be stars.  Group g1
consists of f1 to f3 whereas all six members c1 to c6 of group g2 are
sum or harmonic frequencies of f1 through f3.  The NRP hypothesis
makes it thus possible (but not necessary), by analogy to other Be
stars, to construct c2 and c4 as sum frequencies f1+f2 and f2+f3,
respectively.  This possibility would remove the last deficiency of
the three-separate-frequencies model w.r.t.\, the
single-modulated-frequency hypothesis.  It provides additional
motivation to prefer a triple NRP-mode interpretation for f1 to f3.

Group g0 is empty; this lack of difference frequencies (they would
occur around 0.0021\,c/d) is perhaps not surprising because, in more
active early-type Be stars, they are often associated with repetitive
brightening events that probably signal mass- / angular-momentum-loss
episodes.  $\nu$\,Pup is a very inactive late-type Be star with
mass-loss events possibly occurring as rarely as every couple of
decades.

The NRP diagnosis is consistent with standard properties of NRPs in
B-type stars.  $\nu$\,Pup is located near the cool high-luminosity
'corner' of the OPLIB SPB instability strip calculated by
\citet{2015A&A...580L...9W}.  The by far most detailed study to date
of the photometric variability of a mid-type Be star is that of
KIC\,11971405 by \citet{2017A&A...598A..74P}.  These authors identify
what they call the 'independent pulsation modes' (group g1 in the
above terminology) of KIC\,11971405 in the range 1.3 to 2.3\,c/d.
Considering the difference in spectral type, B7-8\,IIIe ($\nu$\,Pup)
vs.\, B5\,IV-Ve (KIC\,11971405), 0.656\,c/d is a plausible frequency
for NRP modes in $\nu$\,Pup.

The variability of $\nu$\,Pup is somewhat similar to that of
\object{HD\,175869}: This latter star exhibits a single frequency at
0.639\,c/d and a first harmonic with amplitude of more than 50\% of
the base variability.  However, at 0.111 and 0.197\,mmag,
respectively, \citep{2009A&A...506..133G} they are $\sim$2 orders of
magnitude lower than the amplitudes in $\nu$\,Pup.  Higher harmonics
were also reported for HD\,175869 albeit at very low levels.  Because
the observations only extend over 27.3 days, very closely spaced
triple-frequency structures such as in $\nu$\,Pup could not be
detected so that it is not known whether the also reported amplitude
variability is real or caused by a nearby but unresolved frequency.
It also remains unclear what the impact of undetected frequency
multiplicity would be on the seismic modelling performed by
\citet{2012A&A...539A..90N}.

The photometric amplitudes of nonradially pulsating stars depend on
the mode type and inclination angle.  \citet{2003A&A...411..229R}
found that the dominating line-profile variability of most early-type
Be stars is best modeled with retrograde quadrupole NRP modes
($\ell$\,=\,$\vert m \vert$\,=\,2).  Because of the equator-on
perspective implied by the temporary presence of shell lines, such
modes can attain large photometric amplitudes in $\nu$\,Pup
\citep{2007AcA....57...11D}. By contrast, the spectroscopic
sensitivity to such modes is low at inclinations close to 90$^\circ$
\citep{2003A&A...411..229R}, and any multi-mode NRP would further
dilute the spectroscopic signatures.  Therefore, the lack of detection
of the latter (Sect.\,\ref{spectroscopy}) is not a
model-discriminating diagnostic.

An interesting question is how the actually observed frequencies in a
group are selected from all the other possibilities.  It is a variant
of the more general question why from the very dense NRP frequency
spectra of B stars often only a few modes dominate the observed power
spectrum, and which ones.  As Table\,\ref{frqs} proves, the
frequencies in $\nu$\,Pup are extremely tightly coupled.  Either the
strength of this coupling or the strength of the pulsations, or both,
may be responsible for the strong and variable asymmetry of the
individual cycles of the light curve (Figs.\,\ref{LCmulti} and
\ref{BLC}).  It may be frequency differences like 0.0021\,c/d in
$\nu$\,Pup that act as mode couplers and selectors because in some Be
stars half a dozen pairs of apparent NRP modes share the same
difference frequency \citep{2017arXiv170808413B}.  Another
selection/combination rule seems to be that only modes with the same
angular eigenfunction combine in an appreciable way
\citep{1998IAUS..185..347B} in mass-loss events.

The strength of coupling of $g$-modes scales with the square of the
rotation rate and becomes relevant for modes more closely spaced than
the rotational splitting.  The spacing of $\sim$0.0021\,c/d is way
below this threshold.  For the principles and examples of $g$-mode
coupling, see \citet{1998A&A...334..911S} and
\citet{2002A&A...392..151D}, respectively.

Through combination frequencies shared by many pairs of NRP modes,
single NRP frequencies can be assembled to intricate nested clockworks
that open and close the mass- and probably also angular-momentum-loss
valves of Be stars.  A rich example is 25\,Ori
\citep{2017arXiv170808413B}.  If selected difference frequencies act
as mode couplers and selectors also in $\nu$\,Pup, any mass-loss
events from first-level difference frequencies ($\sim$0.0021\,c/d) and
the NRP modes involved in them may be below the detection limits of
SMEI and BRITE.  (A speculative complementary description of the mass
loss from $\nu$\,Pup is outlined in Sect.\,\ref{masslossD}.)
Undetected low-amplitude variabilities may also explain the very low
relative widths of the two frequency groups in $\nu$\,Pup.  At
$\Delta f/f\,\ll$\,1\%, they are narrower by much more than an order
of magnitude than typical frequency groups in other Be stars.  The
narrowness of the groups could be just the result of the close
sub-grouping of these high-amplitude frequencies.

This latter conclusion is supported by the case of the pulsating Be
star 25\,Ori where three similar triple-frequency structures appear in
the power spectrum \citep[][Baade et al., in
prep.]{2017arXiv170808413B}.  All three have the spacing of
0.0129\,c/d in common and, together with additional, weaker features,
form frequency subgroups, which are correspondingly narrow but still
more than six times as wide as the f1 to f3 complex in $\nu$\,Pup.  Two of
them fall into g1, one belongs to g2.  However, there are no
detectable harmonics.

Although 0.0021\,c/d is not detected as a difference frequency, the
clustering near 0.0021\,c/d of frequency differences in $\nu$\,Pup
invites comparisons to accumulations of frequency differences seen in
other Be stars.  In 25\,Ori \citep{2017arXiv170808413B}, at least two
clusters of frequency differences exist, with complex connections and
interactions between them.  Simpler patterns have been found in
\object{$\eta$\,Cen} \citep{2016A&A...588A..56B} and \object{28\,Cyg}
\citep{2018A&A...610A..70B}.  Therefore, this and other mode-selection
rules in Be stars may be the DNA of the stellar part of the Be
phenomenon, and 25\,Ori and $\nu$\,Pup span much of the range of
variability patterns in Be stars.  25\,Ori undergoes cyclicly
repeating outbursts, probably governed by a hierarchical set of
NRP-controlled clocks \citep{2017arXiv170808413B}.  This persistent
activity may lead to the appearance of additional frequencies, which
has been proposed to happen during outbursts
\citep{2009A&A...506...95H, 2018A&A...610A..70B}.  $\nu$\,Pup harbors
a similar clockwork, and the main difference may be that the
observations presented of $\nu$\,Pup depict such a clockwork during
quiescence and not obliterated by the mass-ejection process.

\subsection{Mass loss}
\label{masslossD}
The temporary presence in $\nu$ Pup of line emission and shell
absorption in H$\alpha$, however exiguous they were, requires some
(variable) mass-loss mechanism.  The latter cannot be founded on
rotation or radiation alone or their combination.  A good fraction of
the mass loss from Be stars may take place in discrete events (cf.\
Introduction).  However, both photometry and spectroscopy suggest
that, in $\nu$\,Pup, major events happened at most once per decade.
According to model calculations by \citet{2012ApJ...756..156H},
mass-loss events in equator-on stars like $\nu$\,Pup are effective in
letting the circumstellar disk remove part of the stellar flux.  No
major dimming event was seen in many years of photometry which,
however, had only limited sensitivity to such variations.  There may
be brightenings by a few mmag and for a fraction of a day.  If they do
present mass-loss events, the amounts of matter ejected per event must be 
very small.  But there could be of order 10$^2$ such events per year.  

A rate of one larger outburst or less per decade is within the
findings of \citet{2018MNRAS.479.2909B} for late-type Be stars who, 
in grund-based data, had lower sensitivity to outbursts than this study
of $\nu$\,Pup has.  Without replenishment, the disk is disrupted by
viscous re-accretion as well as viscous expansion into the
interstellar medium \citep{2017MNRAS.464.3071V}.  Given the low UV
flux of B8 stars, additional radiative demolishment \citep[][and
references therein]{2018MNRAS.474..847K} is comparatively unimportant.
Accordingly, unperturbed disk dissipation times can be measured,
providing important constraints on temporal and spatial variations of
the viscosity.  Alternatively, one might have to infer a more
continuous star-to-disk mass-transfer process that is too weak to be
detected by the most common observing techniques.  For experimental
verification in the optical, linear continuum
\citep{1984ApJ...287L..39G, 2007ApJ...671L..49C} and line polarimetry
is more sensitive to near-stellar matter than plain photometry or
spectroscopy are.

If clusters of frequency differences are involved in the mass loss
from $\nu$\,Pup, the only possible choice offered by the observations
is 0.0021\,c/d.  During the time of the observations the corresponding
difference frequencies were not detectable because the conditions for
the driving of mass loss were not met.  If the condition for outbursts
of $\nu$\,Pup is that all of f1 to f3 are in phase, the implied
timescale is 40 years.  This is not too different from the separation
in time of the two shell phases.  The latter interval amounts to 80-90
years (Sect.\,\ref{discussion}), and another event in between would
have gone unnoticed for lack of observations.  $\nu$\,Pup could, then
be another illustration of hierachically nested clocks that drive mass
loss from Be stars \citep{2017arXiv170808413B}.  At the lowest clock
level $\ell$1, there are the NRP modes in g1.  At the third level
$\ell3$, there is the difference between the two difference
frequencies f2-f1 and f3-f1, which both form the second level $\ell$2.
The orders of magnitude of the three time scales are a day, a year,
and some decades, respectively, for $\ell$1, $\ell$2, and $\ell$3.
Apparently, clock level $\ell$2 has no effect on mass loss, certainly
not at a level of magnitude detectable by spectroscopy.  By analogy
to 25\,Ori, the two difference frequencies would attain high
amplitudes only for a few weeks or even less (they are not simple beat
frequencies).  When the two are roughly in phase, the mass-loss valve
opens and the two difference frequencies drive the mass loss.

\subsection{Binarity}
\label{binarityD}

It has been suggested \citep[][and references
therein]{1991A&A...241..419P, 2002A&A...396..937H,
  2018MNRAS.477.5261B} that binarity plays a formative role in the Be
phenomenon.  In fact, Be stars with subluminous companions
\citep{2018ApJ...853..156W} and Be X-ray binaries with compact
companions \citep{2011Ap&SS.332....1R} prove that there are Be stars
that must have gone through a phase of strong interactions with these
companion stars.  Accordingly, many Be stars probably acquired (much
of) their high angular momentum through mass transfer from a another
star.  On the other hand, just one case of an effectively single Be
star without high space velocity would demonstrate that there must be
both double- and single-star evolutionary paths towards the Be
phenomenon.  If Be stars do form in more than one way, it will be be
very stimulating to learn whether the pulsations of Be stars can be
tracers of these formation processes.

In order to consider a Be star effectively single, its separation
should probably be at least one astronomical unit.  The presumably
best-determined mass and orbit of a supposed sdO companion to a Be
star are that of the B2\,Ve star \object{$\phi$\,Per} with
1.2\,M$_{\odot}$ and orbital period of 127\,d
\citep{2015A&A...577A..51M}.  It seems feasible to rule out such a
star in $\nu$\,Pup (Sect.\,\ref{Doppler}), using limits on photometric
Doppler shifts.  With precise modeling and stacking of suitable
features in the power spectrum, even lower detection limits appear
possible.  However, companions with masses as low as 0.07\,M$_{\odot}$
have been reported, based on up to 88 {\it IUE} spectra
\citep[][HR\,2142, B1.5\,IV-Vnne]{2016ApJ...828...47P}.  Such masses
do not seem to be within reach of the present photometry of
$\nu$\,Pup.  Application of the same technique to the {\it IUE}
spectra of the much cooler star $\nu$\,Pup should help.  However, with
just two spectra, comparable sensitivity cannot be achieved.
Nevertheless, $\nu$\,Pup appears as one of the best targets for
efforts aiming at the first identification of an effectively single Be
star.

At 30-35\,km/s, the radial velocity is relatively high; the
plane-of-sky component of $\sim$3\,km/s derived from the Gaia parallax
and proper motion \citep{2018yCat.1345....0G} only adds little to the
space velocity.  This velocity could be the result of the break-up of
a former binary if the more massive star underwent a supernova
explosion.  Since $\nu$\,Pup traverses the current distance of
$\sim$110\,pc from the solar system in about 3.5 10$^6$ years, no
spectacular event would be implied.  However, for a single object,
nothing can be said about any earlier mass and angular-momentum
transfer to the Be star from an assumed exploded star.

\section{Conclusions}

The time spanned - three decades from 1989 through 2018 - and
significantly covered by the space photometry of $\nu$\,Pup is not
exceeded by space observations analyzed for other Be stars.  This has
permitted an unprecedented study of the long-term behavior of the
variability of a Be star.

$\nu$\,Pup is a spectroscopically normal B7-B8\,IIIe star.  It
exhibits a $\sim$10-mmag variability near 0.6556\,c/d split into three
frequencies f1 to f3 spaced by $\sim$0.0021\,c/d.  The frequency
spacing is nearly but not exactly equidistant; neither appear the
periods equidistant.  Five much weaker variabilities form a second
nearly equidistant frequency group near 1.31\,c/d; three of them are
harmonics of f1 to f3.  Similarly narrow groups of high-amplitude NRP
modes are not known from other Be stars unless similar frequency
groups in HD\,175869 were not resolved by the one-month observations
\citep{2009A&A...506..133G}.  However, the narrowness may be the
result of the insensitivity of the observations to sub-mmag
variations.  The frequency spacing is very much smaller than any
expected rotational splitting of NRP modes.

The similarity of the spacing between all frequencies in both groups
suggests that they are somehow coupled.  This coupling along with the
large amplitudes may be responsible for the strong and varying
asymmetry of the light curve.

In early-type Be stars, difference frequencies seem to be involved in
the triggering of mass-loss events.  The existence of very effective
selection rules of NRP modes, which are also implied by other
observations, may be part of the DNA of the stellar component of the
Be phenomenon.  Because the mass-loss process in $\nu$\,Pup is very
weak but frequency combinations are still very prominent, NRP-enabled
mass loss from Be stars should only be the consequence of coupled NRP
modes, not also their cause.  A causal element may lie in the radial
transport of excess angular momentum.  Because Be stars are not
destroyed by the angular momentum rising to the surface, the unknown
mode-selection process may favor difference frequencies effective
in preventing an angular-momentum catastrophe.

Before the availability of space photometry, Be stars were notorious
for their seemingly unpredictable, erratic behavior.  Satellite
observations have revealed complex variability patterns, which can be
governed by multiply nested clocks and, therefore, can be repetitive.
However, the duration of observations required to recognize such
repetitive patterns is long, especially if multiple and/or closely
spaced frequencies rule the light curve.  The 0.0021-c/d frequency
spacing in $\nu$\,Pup has extended the range of these timescales to
$\sim$1000\,d.  The synchronization over two very similar
$\sim$0.0021-c/d beat cycles in $\nu$\,Pup of f1 to f3 takes decades,
possibly in agreement with the spacing of shell episodes.

Late-type Be stars are also a source of valuable observational data
for the analysis of the disk dissipation process: Compared to hotter
Be stars, radiative effects on the disk are much lower, the
dissipation is much less likely to be disturbed or reversed by new
outbursts, and the sensitivity to a continuous star-to-disk
mass-transfer process, if any, could be much higher.  This may enable a more
precise determination of the viscosity parameter, its radial and
vertical distribution in the disk, and its temporary evolution
\citep[cf.\ the work on the early-type Be star 
\object{28\,CMa} by][]{2018MNRAS.479.2214G}.

Motivated by the expectation that Be stars result from the evolution
of binary stars, searches have been carried out, and many actual and
candidate binary systems have been idemtified.  It would be most
valuable to complement such efforts with deep analyses of a few Be
stars with the goal of finding out whether effectively single Be stars
also exist.  The edge-on perspective, relatively low mass, very large
photometric amplitude, and simple frequency spectrum render $\nu$\,Pup
a promising test target.  By the example of $\nu$\,Pup it could be
shown, for the first time in Be stars, that photometric Doppler shifts
can place useful constraints on the properties of companion stars.
Other Be stars worth investigating in this way include KIC\,11971405
and Achernar \citep{2011MNRAS.411..162G}.  If effectively single Be
stars can be identified, it will be most illuminating to learn whether
pulsations can distinguish the formation histories of different
populations of Be stars.

\begin{acknowledgements} The authors are immensily indebted to
  Professor Bernard V.\ Jackson for the provision of the newly
  processed SMEI observations.  They thank the BRITE operations staff
  for their untiring efforts to deliver data of the quality that
  enabled this investigation.  Dr.\ Coralie Neiner is gratefully
  acknowledged for the in-advance communication (through GAW) of the
  results of the magnetic-field measurement.  The Polish contribution
  to the BRITE project is supported by Polish Ministry of Science and
  Higher Education, and the Polish National Science Center (NCN, grant
  2011/01/M/ST9/05914).  This research has made use of the SIMBAD
  database, operated at CDS, Strasbourg, France.  This research has
  made use of NASA's Astrophysics Data System.  This work is partly
  based on observations obtained at the Canada-France-Hawaii Telescope
  (CFHT) which is operated by the National Research Council of Canada,
  the Institut National des Sciences de l’Univers of the Centre
  National de la Recherche Scientifique of France, and the University
  of Hawaii. The CFHT data were obtained from the Canadian Astronomy
  Data Centre operated by the National Research Council of Canada with
  the support of the Canadian Space Agency.  Some of the data
  presented in this paper were obtained from the Mikulski Archive for
  Space Telescopes (MAST). STScI is operated by the Association of
  Universities for Research in Astronomy, Inc., under NASA contract
  NAS5-26555. Support for MAST for non-HST data is provided by the
  NASA Office of Space Science via grant NNX09AF08G and by other
  grants and contracts.  ACC acknowledges support from CNPq
  (grant307594/2015-7).  GH thanks the Polish NCN for support (grant
  2015/18/A/ST9/00578).  AFJM is grateful for financial aid from NSERC
  (Canada) and FRQNT (Quebec).  APi acknowledges support from the
  Polish NCN grant no. 2016/21/B/ST9/01126.  DP acknowledges financial
  support from Conselho Nacional de Desenvolvimento Cient\'ifico e
  Tecnol\'ogico (CNPq-MCTIC Brazil; grant \mbox{300235/2017-8}).  SMR
  and GAW acknowledge Discovery Grant support from the Natural
  Sciences and Engineering Research Council (NSERC) of Canada.  KZ
  acknowledges support by the Austrian Fonds zur F\"orderung der
  wissenschaftlichen Forschung (FWF, project V431-NBL) and by the
  Austrian Forschungsf\"orderungsgesellschaft (FFG).
\end{acknowledgements}

\bibliography{dbaade}

\end{document}